\documentclass[useAMS,usenatbib]{mn2e}
\usepackage{epsfig}
\usepackage{amssymb}
\usepackage{amsmath}
\usepackage{lscape}
\usepackage{longtable}

\def\arcsec{$^{\prime\prime}$}

\title[The central energy source of 70$\mu$m-selected galaxies: Starburst or AGN?]{The
central energy source of 70$\mu$m-selected galaxies: Starburst or AGN?}
\author[Symeonidis et al.]{M. Symeonidis$^{1}$\thanks{msy@mssl.ucl.ac.uk}, D. Rosario $^{2}$,
A. Georgakakis$^{3}$, J. Harker$^{2}$, E. S. Laird$^{4}$ and \and M. J. Page$^{1}$ \\
$^{1}$ Mullard Space Science Laboratory, University College London, Holmbury St. Mary, Dorking,
Surrey RH5 6NT, UK\\
$^{2}$ Astronomy $\&$ Astrophysics, 201 Interdisciplinary Sciences Building,
Santa Cruz, CA 95064, USA\\
$^{3}$ National Observatory of Athens, Institute of Astronomy, V. Paulou
and I. Metaxa, Athens 15236, Greece\\
$^{4}$ Imperial College London, Blackett Laboratory, Prince Consort Road, London SW7 2AZ, UK}

\begin{document}

\date{Accepted  Received; in original form}

\pagerange{\pageref{firstpage}--\pageref{lastpage}} \pubyear{2009}

\maketitle

\label{firstpage}

\begin{abstract}
We present the first AGN census in a sample of 61 galaxies selected at
 70\,$\mu$m, a wavelength which should strongly favour the detection of
 star-forming systems. For the purpose of this study we take advantage
 of deep \textit{Chandra} X-ray and \textit{Spitzer} infrared
 (3.6-160\,$\mu$m) data, as well as optical spectroscopy and photometry
 from the Deep Extragalactic Evolutionary Probe 2 (DEEP\,2)
 survey for the Extended Groth Strip (EGS) field. We investigate spectral line diagnostics ([OIII]/H$\beta$ and [NeIII]/[OII] ratios, H$\delta$
 Balmer absorption line equivalent widths and the strength of the
 4000\,$\AA$ break), X-ray luminosities and spectral energy distributions (SEDs). 
We find that the 70\,$\mu$m sources are undergoing starburst episodes and are therefore characterised by a predominance of young stars. In addition, 13 per cent of the sources show
 AGN signatures and hence potentially host an AGN. When the sample is
 split into starbursts (SBs, 10$^{10}$\,$<$\,L$_{IR}$\,$<$\,10$^{11}$\,L$_{\odot}$), Luminous
 InfraRed Galaxies (LIRGs, 10$^{11}$\,$<$\,L$_{IR}$\,$<$\,10$^{12}$\,L$_{\odot}$) and
 UltraLuminous InfraRed Galaxies (ULIRGs,
 10$^{12}$\,$<$\,L$_{IR}$\,$<$\,10$^{13}$\,L$_{\odot}$), the AGN
 fraction becomes 0, 11 and 23 per cent
 respectively, showing an increase with total infrared
 luminosity. However, by examining the sources' panchromatic SEDs, we conclude that although the AGN is
 energetically important in 1 out of 61 objects, all
 70\,$\mu$m-selected galaxies are primarily powered by
 star-formation. When compared to a sample of DEEP\,2 galaxies in the same redshift range and
 with similar optical colours, we find that the 70\,$\mu$m population is
 characterised by younger stellar ages and a higher AGN incidence,
 indicating that strongly star-forming populations might be key in
 studying the relationship between black hole and stellar growth.

\end{abstract}

\section{Introduction}

Understanding the relationship and relative contribution of the
fundamental energy sources in the Universe, AGN accretion
and star formation, is far from trivial. Data from three
generations of infrared observatories, \textit{IRAS}, \textit{ISO} and \textit{Spitzer}, have revealed that the most
powerful activity in the Universe is dust-enshrouded. Obscured galaxies are responsible for roughly half
the cosmic energy density originating from stars and AGN: the Cosmic InfraRed Background (CIRB). Observational and theoretical results have placed the history of star formation and black hole accretion in parallel evolutionary paths, linked to the close relationship between central black holes and galaxy properties (Magorrian et al. 1998\nocite{M98}), a key aspect of identifying the
processes that drive galaxy evolution (e.g. Sanders et al. 1988\nocite{S88}; Norman $\&$ Scoville 1988\nocite{NS88}; Rowan-Robinson $\&$ Crawford 1989\nocite{RRC89}; Rowan-Robinson 1995\nocite{RR95}), with speculations of a complex two-way feedback method whereby AGN and starburst
activity is inter-regulated (e.g. Umemura, Fukue $\&$ Mineshige 1997\nocite{UFM97}; Appleton et al. 2002\nocite{A02};
King 2005\nocite{K05}; Springel, Di Matteo $\&$ Hernquist 2005\nocite{SdMH05}). 
Although, disentangling the relative starburst/AGN contributions to the
infrared energy budget is key for identifying the processes that drive
galaxy evolution, such a task has proved extremely challenging and any substantial
progress has been slowed by the difficulty of identifying obscured AGN at high
redshift. The presence of dust in the host galaxy complicates matters
further, as the relative contribution of stars and AGN to the bolometric luminosity becomes an elusive quantity. 

The main contributors to the CIRB, Luminous and
UltraLuminous Infrared Galaxies (LIRGs and ULIRGs), are characterised by
large amounts of dust, emission from which completely dominates their energetic output. As dust grain formation and
destruction is a dynamical, stellar-related process with the lifetime of
a typical grain of the order of a few hundred Myr (Draine 2003\nocite{D03}), the
presence of large amounts of dust combined with a high infrared output implies rapid, ongoing
star-formation and/or intense starburst episodes. With data from the Multiband Imaging Photometer for \textit{Spitzer} (MIPS), the properties of
these populations have been examined out to z\,$\sim$\,2, however it is
only with the far-IR (70 and 160\,$\mu$m) bands that greater relevance
to star-formation can be achieved, as they probe nearer the SED peak
of a star-forming galaxy ($\sim$\,40--150\,$\mu$m). Because of its higher
sensitivity, the MIPS 70\,$\mu$m band has been more widely exploited
than the 160\,$\mu$m band, with 70\,$\mu$m populations found to
predominantly consist of infrared-luminous (L$_{IR}>10^{10}$) and dust-rich objects, mainly in the
0.1$\lesssim$z$\lesssim$1.5 redshift range (Symeonidis et al. 2008\nocite{Sym08};
2009\nocite{Sym09}). As a result, setting the initial sample selection at 70\,$\mu$m is ideal
for studying the relationship (if any) between obscured star-formation and supermassive
black hole growth, as well as evaluating the degree of
coeval activity between AGN and star-formation during a galaxy's lifetime.  

Numerous studies have revealed that a large
part of the bolometric luminosity in IR-luminous sources could originate from AGN in some cases prevailing over the starburst
output (e.g. Gregorich et al. 1995\nocite{G95}; Genzel et al. 1998\nocite{G98}; Tacconi et
al. 2002\nocite{T02}), with the AGN contribution seen to
increase as a function of total infrared
luminosity (e.g. Lutz et al. 1998\nocite{Lutz98}; Fadda et al. 2002\nocite{Fadda02}; Brand et al. 2006\nocite{Brand06}). Identifying the presence
of AGN in obscured galaxies requires the combination of various disgnostics such as mid-IR colours (e.g. Lacy et al. 2004\nocite{L04}; Stern et
al. 2005\nocite{Stern05}; Barmby et al. 2006\nocite{Barmby06}), SED shape (e.g. Alonso-Herrero et al. 2006\nocite{AH06a}), X-ray luminosity (e.g. Pompilio, La Franca $\&$ Matt 2000\nocite{PLM00}) and
radio luminosity (e.g. Donley et al. 2005\nocite{Donley05}). Optical spectrosopy is also a powerful tool (e.g. Osterbrock 1989\nocite{O89}; Tresse et al. 1996\nocite{T96}; Takata et al. 2006\nocite{T06}), with recent evidence pointing towards
evolutionary differences, rather than solely orientation effects (e.g. Antonucci 1993\nocite{A93}), being responsible for the variety of AGN
spectroscopic signatures (e.g. Hasinger
2008\nocite{H08}), resulting in a mixture of both type I
and type II spectra. In addition to the traditional `torus-obscured' AGN, studies have
revealed the potential of `host-obscured' AGN as a plausible scenario, where
light from the nucleus is also obscured by the dusty star-forming host galaxy
(e.g. Martinez-Sansigre et al. 2006\nocite{MS06}). For such sources, optical spectroscopy is a powerful diagnostic with a plethora of emission lines both from the AGN and host, providing important clues on the environments in which they originate (e.g. Brand et al. 2007\nocite{Brand07}). 70\,$\mu$m populations are at a considerable advantage for multiwavelength spectroscopic and photometric follow-up: the sensitivity of the MIPS 70\,$\mu$m band is such that
only the brightest sources are selected at each redshift and hence are very likely to have counterparts over a wide wavelength range. Moreover, the redshift distribution of 70\,$\mu$m populations allows many key emission and absorption lines to be within a typical optical grating range. 

Due to the effects of dust obscuration, AGN censuses in infrared and sub-mm
populations have mainly been approached through mid-IR and X-ray surveys
(e.g.  Vignati et al. 1999\nocite{Vignati99}; Almaini, Lawrence $\&$ Boyle 1999\nocite{ALB99}; Fabian, Wilman $\&$ Crawford
2002\nocite{FWC02}; Alexander et al. 2005\nocite{A05}; Martinez-Sansigre et al. 2005\nocite{MS05}; Matute et al. 2006\nocite{Matute06}).
X-ray surveys have resolved most of the 0.5--10\,keV Cosmic X-Ray Background (CXRB) and attributed it to emission from accreting super-massive black holes (e.g. Shanks et al. 1991\nocite{S91}), however lack of sensitivity has not allowed the detection of the majority of
extragalactic X-ray sources contributing at low ($<$\,10$^{-16}$\,erg/s/cm$^2$) fluxes. These are most likely high-redshift, star-forming and starburst galaxies, possibly containing low luminosity or deeply obscured AGN (e.g. Ptak et al. 2003\nocite{P03}; Franceschini et al. 2003\nocite{F03}; Bauer et al. 2004\nocite{Bauer04}; Brandt $\&$ Hasinger 2005\nocite{BH05}). 
A large obscured AGN population has long been predicted
both from statistical analysis and number counts of AGN samples as well
as from CXRB models (e.g. Comastri et
al. 1995\nocite{C95}; Worsley et al. 2005\nocite{W05}), especially since the steep broad-line
QSO spectrum cannot explain the intensity of the hard X-ray background (e.g. Fabian et
al. 1998\nocite{F98}). This implies that most accretion-generated energy density in the
Universe takes place in obscured objects, making obscured AGN the main candidates for the origin of the hard CXRB, which peaks at
$\sim$30\,keV and, as yet, has not been resolved (e.g. Setti $\&$
Woltjer 1989\nocite{SW89}; Fabian $\&$ Iwasawa
1999\nocite{FI99}).

The aim of this paper is to examine the AGN content of a sample of IR-luminous 70\,$\mu$m-selected galaxies from the Extended Groth Strip (EGS)
field. We combine \textit{Chandra} X-ray data and optical spectroscopy in
order to constrain the AGN fraction and identify the dominant process responsible for the sources' total energy budget. Section 2 introduces the sample and gives an analysis of IR/optical colours. In section 3 we examine the sample's X-ray properties, whereas in section 4 the analysis focuses on optical spectra and emission line ratios. The AGN contribution is discussed in section 5 and our summary and conclusions are presented in section 6. 
Throughout we employ $H_0=70$ kms$^{-1}$Mpc$^{-1}$, $\Omega_M=0.3$ and
$\Omega_{\Lambda}=0.7$ (Spergel et al. 2003\nocite{S03}).

\section{The Sample}

\subsection{Initial selection and previous work on the 70\,$\mu$m sample}
\label{sec:previous_work}

This work is based on Guaranteed Time Observations (GTO) of the Extended Groth Strip (EGS) field ($\sim$0.5\,deg$^2$) (Davis et al. 2007\nocite{D07}) by \textit{Spitzer's} (Werner et al. 2004\nocite{W04}) far-IR photometer MIPS (Rieke et al. 2004\nocite{Rieke04}). The initial selection was made at 70\,$\mu$m, where 178 sources were retrieved down to $\sim$\,4\,mJy (5$\sigma$), with photometric completeness at $\sim$\,10\,mJy. We focus on a subset of this population, the 114 sources in the overlap areas of the MIPS 24, 70, 160\,$\mu$m and InfraRed Array Camera (IRAC) 8\,$\mu$m images and further narrow our working sample to 61 sources with optical photometry and spectroscopic redshifts from the Deep Extragalactic Evolutionary Probe 2 (DEEP\,2) survey (see section \ref{sec:optical_selection}). The MIPS 70\,$\mu$m catalog was cross-correlated to the MIPS 24\,$\mu$m and IRAC catalogs in two steps: first using a $\sim$\,10\,\arcsec radius for the MIPS70/MIPS24 matching and subsequently using a $\sim$\,2.5\,\arcsec
radius for the MIPS24/IRAC matching. 
For data reduction and source extraction we refer the reader to Symeonidis et al. (2007\nocite{Sym07}; 2008\nocite{Sym08}; 2009\nocite{Sym09}, hereafter S07; S08; S09), where the infrared properties of the 70\,$\mu$m sample were also examined in detail. A succinct summary of our results from previous work follows:

In S08 and S09, we fitted the available IR photometry (8--160\,$\mu$m) with the Siebenmorgen $\&$ Krugel (2007\nocite{SK07}) templates and obtained estimates for the total infrared luminosity (L$_{IR}$) in the 8--1000\,$\mu$m range. We found characteristic luminosities of IR-luminous galaxies between 10$^{10}$ and 10$^{14}$ L$_{\odot}$, of which $\sim$11$\%$ are Starbursts (SBs, 10$^{10}$ $<$ L$_{IR}$ $<$ 10$^{11}$),
$\sim$62$\%$ are Luminous InfraRed Galaxies (LIRGs, 10$^{11}$ $<$ L$_{IR}$ $<$ 10$^{12}$) and
$\sim$26$\%$ are UltraLuminous InfraRed Galaxies (ULIRGs, 10$^{12}$ $<$ L$_{IR}$ $<$ 10$^{13}$). We also
identified one HyperLuminous InfraRed Galaxy (HyLIRG, L$_{IR}$ $>$ 10$^{13}$) at
z=1.9, which we remove from subsequent analysis as it is unique and will be treated separately in a later paper. The optical/IR SEDs for all, but one, objects are starburst-type, with a strong optical/near-IR stellar bump, an inflection in the near-IR and elevated flux in the infrared. 
One object, identified with red near-IR colours in all IRAC colour bands, has an SED that
shows an elevated near-IR power-law continuum instead of an inflection
in that region (see sections \ref{sec:agn} and \ref{sec:fraction}). Infrared and radio derived star formation rates (SFRs) were calculated in S07 and were found to be in the $\sim$\,5--1000\,M$_{\odot}$/yr range, with the majority of the objects having SFR\,$>$\,50\,M$_{\odot}$/yr. SFRs calculated using optical emission lines not corrected for dust extinction, on average fell short by a factor of 50, implying that the optical part of the spectrum is not an accurate tracer of star-formation in these systems, especially since in addition to high extinction, IR-luminous sources will likely host deeply embedded star-forming regions that do not contribute at all to the optical energy budget.

\begin{figure}
\epsfig{file=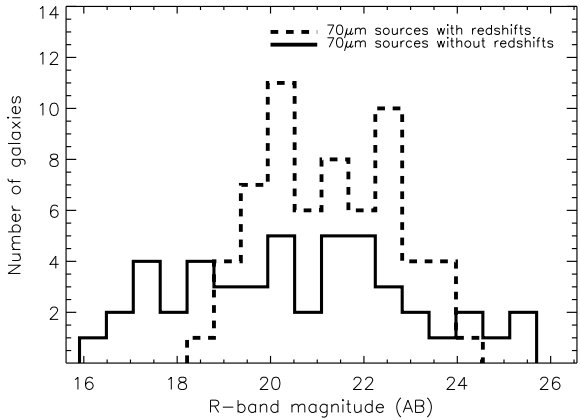,width=8.7cm}
\caption{The R-band magnitude distribution for the 109 70\,$\mu$m sources --- solid line: the 47 sources without redshifts; dashed line: the 61 objects with DEEP\,2
 redshifts. The exclusion of the brightest sources and spectroscopic weighting scheme, cause discrepancies in the two distributions at the faint and bright ends.}
\label{fig:Rmagdist}
\end{figure}

\begin{figure}
\epsfig{file=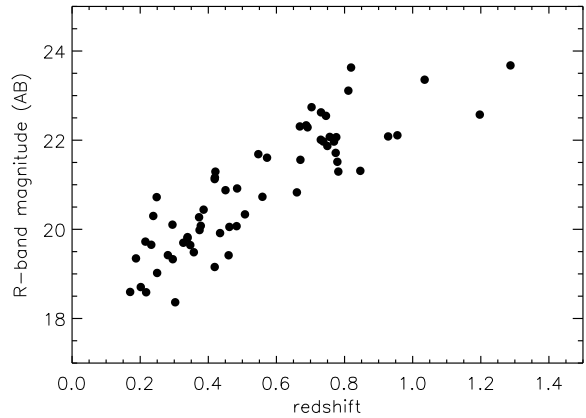,width=8.7cm}
\caption{R-band magnitude versus redshift.}
\label{fig:Rmag_z}
\end{figure}

\begin{figure}
\epsfig{file=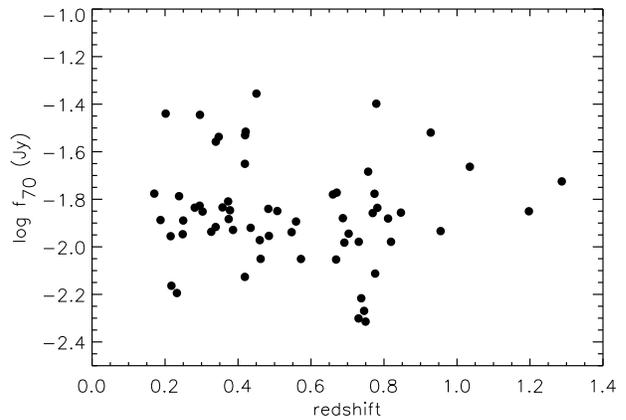,width=8.7cm}
\caption{70\,$\mu$m flux density versus redshift.}
\label{fig:R70_z}
\end{figure}

\begin{figure*}
\epsfig{file=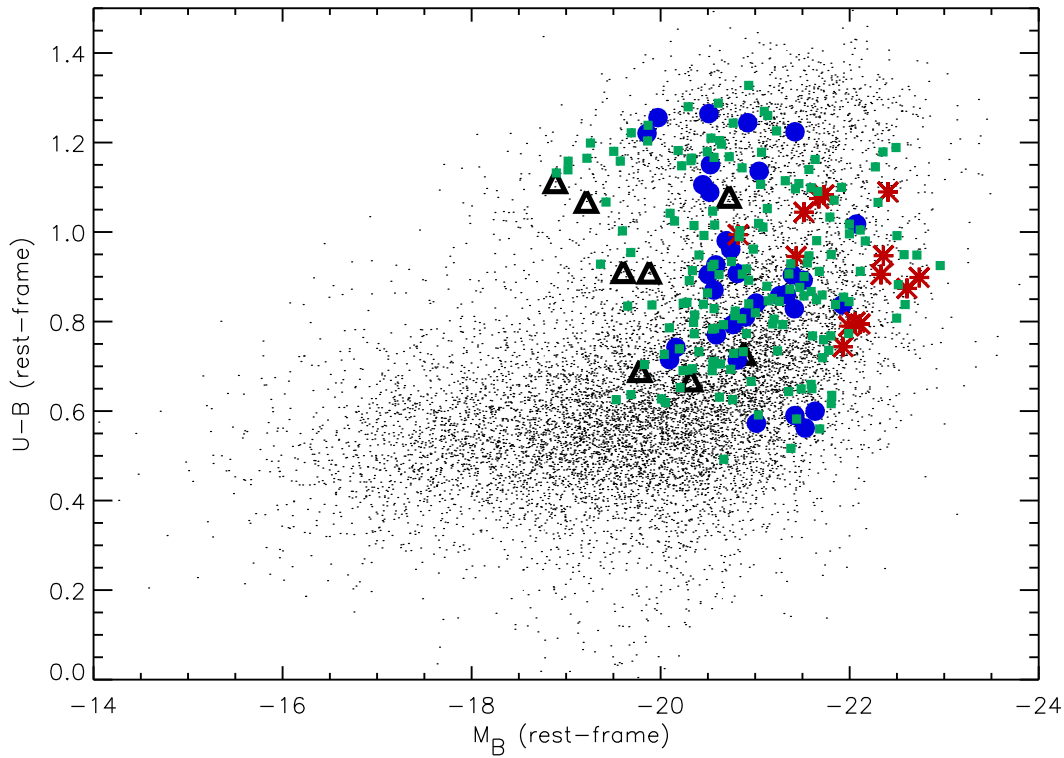,width=15cm}
\caption{U-B colours vs absolute B magnitude, M$_B$ (AB), with
the 70\,$\mu$m sample in black open triangles, blue filled circles and red asterisks(SBs, LIRGs and ULIRGs, respectively) and the control sample as
green filled squares. All $\sim$\,11,000 EGS source from the DEEP\,2 spectroscopic survey are shown in the background as black points.}
\label{fig:CMD}
\end{figure*}

\subsection{Optical selection}
\label{sec:optical_selection}

The EGS benefits from extensive observational coverage via the All-Wavelength Extended Groth Strip International Survey (AEGIS), which provides optical spectroscopy and multi-wavelength photometry over an area of $\sim$\,1\,deg$^2$ (see Davis et al. 2007\nocite{D07} for a summary of the AEGIS photometric datasets). 
From the 114 objects with infrared photometry, we select 109 with area coverage in the B, R, I bands, from the DEEP\,2 Survey  (Coil et al. 2006\nocite{Coil06}; Faber et al. in prep.). Taking advantage of IRAC's high positional accuracy ($\sim$0.1\arcsec), we cross-correlate the 70\,$\mu$m sample with DEEP\,2 using the IRAC coordinates for our sources and an appropriate matching radius of $\lesssim$\,1.5\,\arcsec corresponding to IRAC's pixel size. Currently, $\sim$\,75 per cent  of the MIPS 70\,$\mu$m strip is covered by DEEP\,2 and this will increase to 100 per cent with the DEEP\,3 cycle of observations. Where possible, we also supplement our dataset with photometry from the Galaxy Evolution Explorer (GALEX), the Palomar Wide-field InfraRed Camera (WIRC) and the Canada-France-Hawaii Telescope Legacy Survey (CFHTLS) in the FUV, NUV, g, r, i, z, J and K bands and \textit{Spitzer's} IRAC in the 3.6, 4.5 and 5.8\,$\mu$m bands. 

Apart from the exclusion of stars, the photometric part of DEEP\,2 did not pose any constraints on the EGS. On the other hand, the spectroscopic part of DEEP\,2 targeted sources only within the 18.5\,$<$\,R\,$<$\,24.1 (AB) magnitude range. Although the retrieval of redshifts was random at $\sim$60 per cent sampling, a weighting scheme was applied, giving lower weights to z$<$0.75 objects in order to sample a wide range of luminosities and roughly equal numbers of galaxies below and above z=0.75; see Faber et al. (2007\nocite{F07}); Davis et al. (2007\nocite{D07}). For a reliable redshift, two significant features, such as the [OII] doublet, were required in order to provide a match and hence most redshift failures were from high-redshift objects (z\,$>$\,1.4) which did not have strong features in the DEEP\,2 spectral range. At the time of writing there are 62 good quality redshifts and spectra available for the 70\,$\mu$m EGS population, in the 0.17\,$<$\,z\,$<$\,1.9 range, with a mean of 0.57 and a median of 0.48. Note that we remove the HyLIRG from the redshift sample (see section \ref{sec:previous_work}) with the final number of sources being 61. Hereafter the term `70\,$\mu$m sample' refers only to the sources with redshifts. 

In figure \ref{fig:Rmagdist} we compare the R-band magnitude distribution of the redshift sample to the sources with no redshifts. Although there is high overall consistency, a few differences are evident originating from the spectroscopic selection criteria outlined earlier. Firstly, there is an under-representation of bright sources: there are 10 R\,$<$18.5 sources in total and, at 60 per cent redshift sampling, $\sim$\,6 of those should have made it in the redshift sample. The R-band -- redshift relation (Fig. \ref{fig:Rmag_z}) shows that these are at low redshift. Secondly, on the faint end, there are 4 objects fainter than R\,=24.1 in the full sample, i.e. 2-3 should have also appeared in the redshift sample. The effects of the spectroscopic selection are further examined in figure \ref{fig:R70_z}, where the 70\,$\mu$m flux density (f$_{70}$) is plotted as a function of redshift. In contrast to the clear R-z relation (Fig. \ref{fig:Rmag_z}), f$_{70}$ does not show signs of a trend with redshift, implying that the DEEP\,2 spectroscopic criteria do not have any affect on the infrared properties of the 70\,$\mu$m sample.

\begin{figure}
\epsfig{file=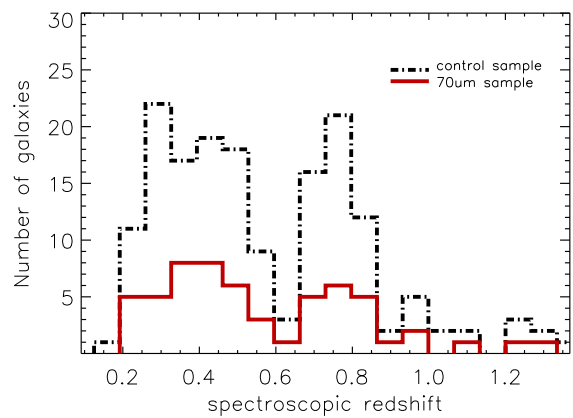,width=8.7cm}
\caption{The redshift distribution of the 70\,$\mu$m sample (red solid
 line) and the control sample (black dashed line)}
\label{fig:redshifts}
\end{figure}

\subsection{The Control Sample}
\label{sec:control_sample}

Due to the unique nature of the 70\,$\mu$m population (see section \ref{sec:previous_work}) and as this work focuses on determining the AGN fraction, it is more insightful to do so in relation to galaxies of comparable properties in a similar redshift range. With that in mind, we assemble a control sample of DEEP\,2 EGS sources: for each 70\,$\mu$m galaxy with a DEEP\,2 redshift, we find up to three others with comparable redshifts, rest-frame B-band luminosities and U-B colours (see Willmer et al. 2006\nocite{Willmer06} for a discussion of \textit{K}-corrections) --- `comparable' is quantified as $\Delta$z\,$\leq$0.03, $\Delta$M$_B$\,$\leq$0.4 and $\Delta$(U-B)\,$\leq$0.1 (see figures \ref{fig:CMD} and \ref{fig:redshifts}). 

Figure \ref{fig:CMD} shows the variation of U-B colour versus M$_{B}$
(AB) --- the Colour Magnitude Diagram (CMD) --- for the control and 70\,$\mu$m sample; the colour distribution for all
EGS sources in the DEEP\,2 spectroscopic survey is also shown in the background. The CMD
represents galaxy evolution from the blue cloud --- young stellar
content --- to the red sequence --- evolved stellar populations ---
associated with morphological changes from spirals to ellipticals and
mediated by a transitional green valley stage (e.g. Bell et
al. 2004\nocite{Bell04}). The 70\,$\mu$m sample is bright in the optical
but has red U-B colours; there is substantial contribution to the red
sequence, but more sources seem to be in the upper blue cloud/green
valley region. 

An alternative way to select the control sample could be by stellar mass. However, due to lack of
complete optical/near-IR datasets for the EGS, mass information for the
DEEP\,2 galaxies (see Conselice et
al. 2007) is only available for about 64 per cent of the 70\,$\mu$m
population. The subset of 70\,$\mu$m objects with available stellar masses are characterised by a narrower distribution of higher average mass ($<$log M$>$ =10.87), compared to the control
galaxies ($<$log M$>$ =10.23) (C. Conselice, private communication). The stellar mass distribution of the 70\,$\mu$m galaxies is consistent with previous studies of infrared
galaxies (e.g. Caputi et al. 2006a\nocite{C06a}; 2006b\nocite{C06b}) and comparable to those of
early-types in the z\,$<$\,1 redshift range (e.g. Ferreras et al. 2009\nocite{Ferreras09}). 
For our purpose, colour/magnitude selection of the control sample is more appropriate than stellar mass, because i) stellar masses are not available for the whole 70\,$\mu$m sample and ii) colour/magnitude selection avoids a control sample dominated by high mass
evolved galaxies with little or no star-formation.

Note that the
optical luminosities and colours are not extinction-corrected, which
might imply that the intrinsic properties of the 70\,$\mu$m sample
could be better matched to objects in the more optically luminous part
of the CMD or those whose red colours are principally a consequence of
dust-reddening rather than stellar age. Notwithstanding, the lack of 
correlation in Fig. \ref{fig:CMD} between U-B colour and infrared luminosity for the
70\,$\mu$m sample, implies that subsequent comparisons of each IR
luminosity class with the respective control galaxies will be fair,
since the latter span a large part of the CMD and hence a large range
in properties.

\begin{table*}
\begin{minipage}{126mm}
\centering
\caption{Table of stacked fluxes (erg/s/cm$^2$) and hardness ratios for the 70\,$\mu$m sample split into
 3 luminosity classes (starbursts, LIRGs and ULIRGs; see section \ref{sec:previous_work} for details on the sources' infrared properties). The second column indicates how many objects were
 used in the stacking --- some were removed either because they were too close to a source which could have contaminated the signal or they were associated with an X-ray source. The SB group does not have a formal detection in the hard band, so we quote the 3$\sigma$ upper limit. The hardness ratio is derived using the counts in the soft band and 3$\sigma$ upper limit counts in the hard band.}
\begin{tabular}{|c|c|c|c|c|}
Class & Objects used & f$_{0.5-2}$ & f$_{2-10}$ & HR \\
\hline 
Starbursts & 8 out of 9 & 8.67E-17 (8$\sigma$) & 2.51E-16 (3$\sigma$ upper limit) &-0.3 (upper limit)\\
LIRGs & 29 out of 35 & 5.04E-17 (9$\sigma$)& 1.6E-16 (4$\sigma$)&-0.2\\
ULIRGs & 9 out of 17&1.11E-16 (12$\sigma$)&1.84E-16 (3$\sigma$)&-0.5\\
\end{tabular}
\label{stacking}
\end{minipage}
\end{table*}

\subsection{X-ray Data}
\label{sec:xray_data}

The EGS was surveyed with the \textit{Chandra} Advanced CCD Imaging
Spectrometer (ACIS) to a depth of 200\,ks, over an area of
0.67\,deg$^2$ (Georgakakis et al. 2006\nocite{G06}; Davis et
al. 2007\nocite{D07}; Laird et al. 2009\nocite{Laird09}). The AEGIS X-ray survey and final data products are made available in Laird et al. (2009, L09). L09 describes the cross-matching between the DEEP\,2 and X-ray AEGIS catalogs; we identify the X-ray counterparts of our 70\,$\mu$m sample using DEEP\,2 coordinates which were in turn matched to the IRAC coordinates of our sources (see section \ref{sec:optical_selection}).  We use X-ray data in the soft (0.5-2\,keV) and hard (2-10\,keV) bands (flux limit of $1.1\times10^{-16}$ and $8.2\times10^{-16}$ erg/s/cm$^2$ respectively), choosing fluxes calculated using a Bayesian method, more reliable for faint sources as it corrects for
the Eddington bias --- for details on source extraction and flux estimation see L09. 

Although the Chandra map covers the whole of the MIPS 70\,$\mu$m
strip, only about 16 per cent of the sources have individual X-ray detections (see table \ref{table1}). Estimates for the hardness ratio, calculated as $\rm HR=\frac{H-S}{H+S}$, where H and S refer to the hard (2-7keV) and soft (0.5-2keV) count rates
(counts/s/cm$^2$), are also shown. If there is only a detection in one band, HR is calculated using the
3$\sigma$ upper limits. From the 61 sources in the sample, only 10 have a confirmed detection in the full band (0.5-8\,kev), 2 of which are not detected in any further bands, 4 have detections only in the soft band, 3 in both the soft and hard bands and 1 in only the hard
band. For the remaining objects we performed stacking (see table \ref{stacking}), after splitting the sample into 3 groups according to the total infrared luminosity measurements. The \textit{K}-corrected X-ray luminosities are calculated by assuming a photon index of $\Gamma$\,=\,1.9 and the stacked fluxes are converted to
luminosity using the mean redshift of the group, 0.22 for SBs, 0.5 for
LIRGs and 0.87 for ULIRGs.

\begin{figure*}
\epsfig{file=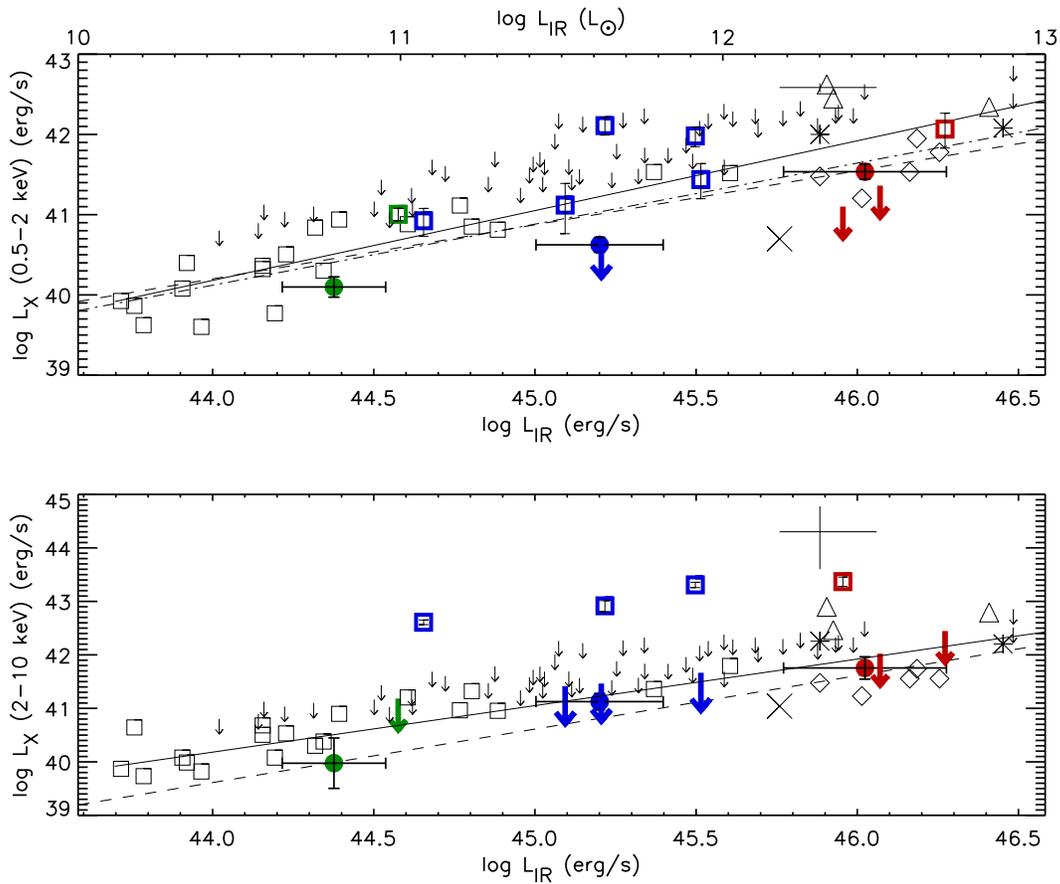,width=15cm}
\caption{Soft (top panel) and hard (lower panel) X-ray luminosity (erg/s) vs $L_{IR}$ (erg/s lower x-axis and L$_{\odot}$ upper x-axis) for the 70\,$\mu$m sample. The colour scheme is green for SBs, blue for LIRGs and red for ULIRGs. Objects with a detection in at least one band have more reliable 3$\sigma$ upper limits in other bands (large coloured symbols), whereas for the remaining sources the upper limits are calculated from the survey flux limits (see section \ref{sec:xray_data}, black symbols). Filled circles correspond to luminosities calculated using the stacked 0.5--2\,keV and 2--10\,keV band flux and open squares are objects with X-ray detections. In the top panel, the lines are the star-forming relations from Griffiths and Padovani (1990) for starburst/interacting galaxies (dashed line) and IRAS galaxies (dot-dashed line) and from Ranalli, Comastri $\&$ Setti (2003) for HII galaxies (solid line). In the lower panel the lines indicate the star-forming relations from Franceschini et al. 2003 (dashed line) for a sample of ULIRGs and Ranalli, Comastri $\&$ Setti (2003) for HII galaxies (solid line). Overplotted is the ULIRG sample from Franceschini et al. (2003); ULIRGs which are SB-dominated in the X-rays (diamonds), ULIRGs which are AGN-dominated in the X-rays (triangles) and ULIRGs which are SB/AGN hybrids in the X-rays (asterisks). The open squares are the HII galaxies from Ranalli, Comastri $\&$ Setti (2003). The small cross is Arp220 and the large vertical cross is NGC6240, where the extent of the cross represents uncertainties in the luminosity.} 
\label{fig:xrayIR}
\end{figure*}

\section{The X-ray/IR relation}
\label{sec:IR}

Figure \ref{fig:xrayIR} compares X-ray and infrared luminosities for the 70\,$\mu$m sample. Given that there are several sources of X-ray luminosity associated with stellar-related processes (e.g. X-ray binaries, supernova remnants, hot gas, starburst-driven outflows), X-ray and infrared emission are expected to be strongly correlated in star-forming galaxies  (e.g. Fabbiano 1989\nocite{F89}; Griffiths $\&$ Padovani 1990\nocite{GP90}; Ranalli, Comastri $\&$ Setti 2003\nocite{RCS03}; Franceschini et al. 2003\nocite{F03}), for which L$_{0.5-10keV}$ ranges between 10$^{40}$--10$^{42}$erg/s (Kim, Fabbiano
$\&$ Trinchieri 1992a\nocite{KFT92a}; 1992b\nocite{KFT92b}; Nandra et al. 2002\nocite{N02}; Laird et al. 2005\nocite{Laird05}).  The boundaries between star-formation and accretion typically extend over a couple of orders of magnitude, nonetheless a hard X-ray luminosity of L$_{2-10keV}$\,$>$\,10$^{42}$erg/s is thought to be an indication that an AGN is the dominant emitter of X-rays (e.g. Georgakakis et al. 2007\nocite{Georgakakis07}). 

We evaluate the properties of our sources against the empirical soft and hard X-ray/IR relations derived for \textit{IRAS} galaxies in the 9$<$log\,L$_{60}$\,$<$\,11.5 range from Griffiths $\&$ Padovani (1990\nocite{GP90}) and for local star-forming SBs and LIRGs from Ranalli, Comastri $\&$ Setti (2003\nocite{RCS03}, hereafter RCS03), as well as the hard X-ray/IR relation derived for a sample of local ULIRGs from Franceschini et al. (2003, hereafter F03). From figure \ref{fig:xrayIR}, it is clear that the survey limits are displaced from the star-forming relations by up to 2 orders of magnitude, suggesting that a 200\,ks X-ray survey is not on average sensitive to distant star-forming galaxies and hence more likely to detect objects with a strong AGN contribution to their X-ray luminosity. 

In the soft X-rays (Fig. \ref{fig:xrayIR}, top panel), the AGN/star-formation separation is not clear: there is large scatter in L$_{0.5-2\rm keV}$, partly due to the fact that soft X-rays are subject to stronger attenuation by gas and dust, the degree of which is likely to vary from one system to another and as a result the contribution of the AGN to the soft X-rays may not be evident. In fact, even some of the ULIRGs in the F03 sample whose X-ray emission is classified as AGN-dominated by F03 seem consistent with the star-formation relations. On the other hand, the 4 sources detected in the hard X-rays (objects 59, 67, 83 and 122, Fig. \ref{fig:xrayIR}, lower panel) have luminosities of L$_{2-10keV}$\,$>$10$^{42}$\,erg/s, 2-3 orders of magnitude higher than expected from the IR-X-ray relation for star-forming systems, placing them into the AGN regime. Their luminosities lie between the hybrid/AGN F03 ULIRG sample and NGC6240 (Vignati et al. 1999\nocite{Vignati99}), in which the X-ray emission is associated with a binary AGN system. The hard X-ray luminosities of the remaining sources are consistent both with the RCS03 and F03 starburst galaxies, as well as with the starburst-dominated local ULIRG Arp220 (Iwasawa et al. 2001\nocite{Iwasawa01}).

\begin{figure*}
\epsfig{file=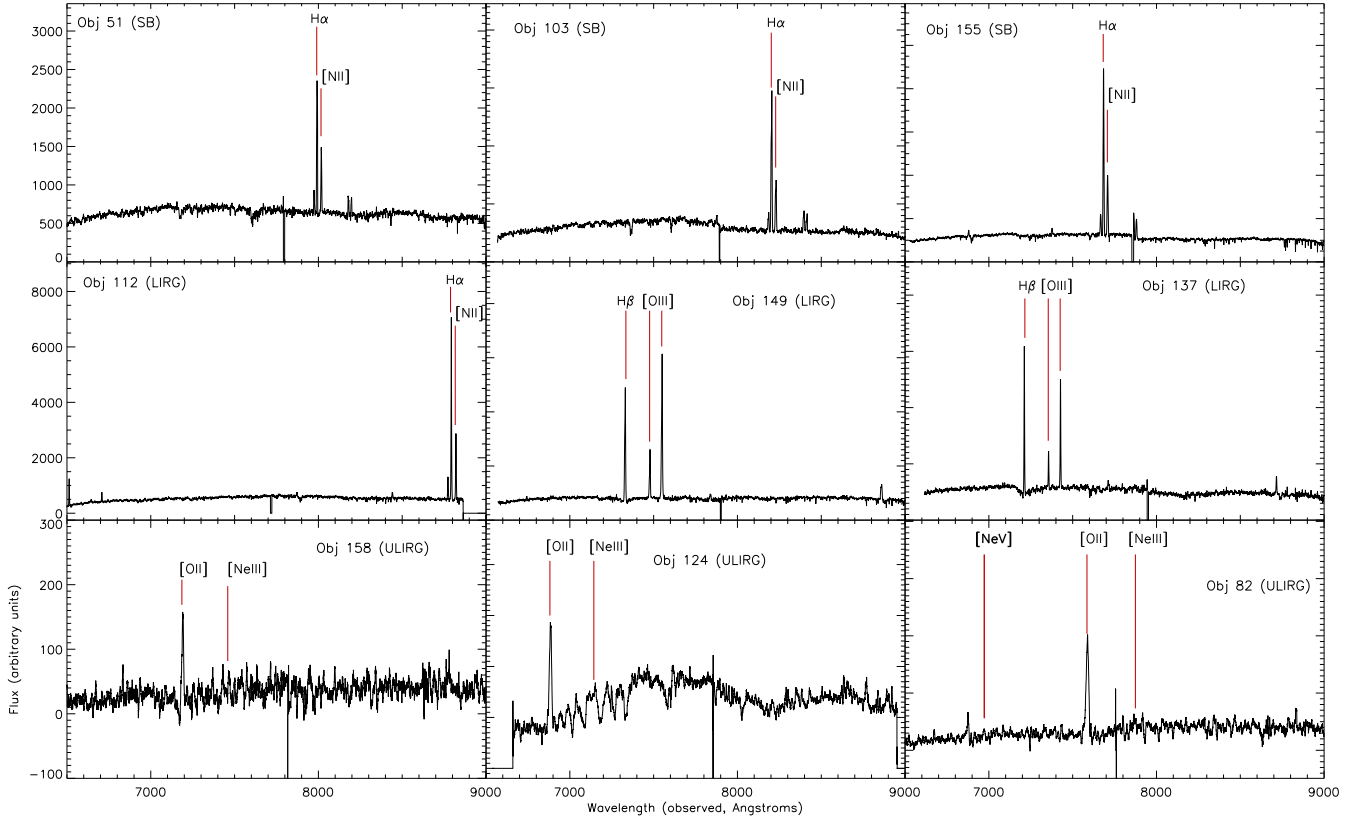,width=18cm}
\caption{Characteristic spectra for the 70\,$\mu$m sample, 3 SBs,
 3 LIRGs and 3 ULIRGs.} 
\label{fig:spectra_all}
\end{figure*}

\section{Optical spectra}
\label{sec:spectra}

Figure \ref{fig:spectra_all} shows characteristic spectra for objects in
the SB, LIRG and ULIRG luminosity classes ($<$\,z\,$>$\,$\sim$\,0.2, 0.5 and 0.8, respectively). The DEEP\,2 spectral range, $\sim$\,6500--9000\,$\rm \AA$, permits the following principal lines to be visible: [NeIII]($\lambda3869$), [OII]
($\lambda3727$) and H$\delta$($\lambda4102$) at 0.6\,$\lesssim$\,z$\lesssim$\,1.8, H$\alpha$ ($\lambda6563$) and [NII] ($\lambda6584$) at z\,$\lesssim$\,0.3, H$\beta$($\lambda4861$) and  the [OIII] doublet($\lambda \lambda 4959, 5007$) at 0.3\,$\lesssim$\,z$\lesssim$\,0.8. Emission line measurements (table \ref {table1}) were made by using single Gaussian fits, after fitting the local stellar continuum with a linear combination of early-type and A-star spectral templates, to approximate the effects of absorption and continuum shape on the line strengths. This is particularly important for faint lines, such as [NeIII], which lie in spectral regions with strong continuum sub-structure.
H$\delta$ equivalent widths (EWs) were derived using the model-fitting method of Harker et al. (in prep.): for the Balmer lines, the emission component, which comes from HII regions, follows a well-defined decrement dependent only on the extinction in the ionized gas. The EWs of the absorption line component, determined by the mean stellar populations, are also well-behaved across a range of star formation histories. By simultaneously measuring H$\beta$ and H$\delta$ in a spectrum, we are able to self-consistently estimate both emission and absorption line EWs for the z\,$\sim$\,0.6-0.8 subset of 70\,$\mu$m and control galaxies. In a similar way, H$\beta$ emission lines were corrected for the underlying H$\beta$ absorption component.

The diagnostic properties of emission lines, particularly useful when expressed in terms of line ratios, have found extensive use in determining conditions in the ISM, stellar ages, metallicity and the strength of the ionisation field. As means of separating nuclear and stellar excitation processes, one of the most widely-used techniques, proposed by Baldwin, Phillips $\&$ Terlevich (BPT, 1981\nocite{BPT81}), relies on identifying sources with respect to their [OIII]/H\,$\beta$ and [NII]/H\,$\alpha$ ratios. As our sample spans a large redshift range, no emission lines are common to the entire sample and as a result we cannot take advantage of such a diagnostic; there are no spectra with all 4 lines simultaneously visible. H$\alpha$ is prominent in the $<$\,z\,$>$\,$\sim$\,0.2 SB population, also seen in the low redshift LIRGs, but redshifted out of the spectral range at z\,$>$\,0.35, where it gets replaced by emission from H$\beta$. The strong H$\alpha$ and H$\beta$ signatures in the galaxies' spectra are indicative of rapid ongoing star-formation as they are directly related to the flux of massive stars ionising the ISM (e.g. Kewley, Geller $\&$ Jansen 2004\nocite{KGJ04}). We find no indication of Doppler broadened (1000\,$\lesssim$\,FWHM\,$\lesssim$\,10000\,kms$^{-1}$) hydrogen lines suggestive of the presence of the broad line region (BLR) around an active nucleus. All sources with H$\alpha$ in the spectral range also show the [NII] feature, however the latter is significantly weaker (e.g. see Fig. \ref{fig:spectra_all}). 
We also identify the 4000\,$\rm \AA$ spectral break (D4000) and the higher order Balmer absorption line H$\delta$, associated with the age of stellar populations, in particular the presence of BAF class stars. D4000 was defined by Bruzual (1983\nocite{Bruzual83}) as the ratio of the average flux density in the 4050-4250\,$\rm \AA$ and 3750-3950\,$\rm \AA$ band, with the narrower equivalent (D$_n$4000: 4000-4100\,$\rm \AA$ to 3850-3950\,$\rm \AA$) defined by Balogh et al. (1999\nocite{B99}). The existence of a break manifests a change in stellar opacity due to the accumulation of a large number of absorption lines from multiply ionised metals in the atmospheres of hot, metal-rich stars. Here we calculate D$_n$4000 as it is less sensitive to extinction, hence more appropriate for our sample and investigate its strength in relation to the Balmer absorption line H\,$\delta$, both visible in the $\sim$\,0.7--1.2 redshift range. However, as calculating H$\delta$ EWs requires reliable H$\beta$ measurements, we only have H$\delta$ absorption line information for the 8 objects in the sample for which both lines are in the accessible range (6 LIRGs and 2 ULIRGs). Moreover, we are able to determine D$_n$\,4000 for only 7 out of 22 sources. For the remaining objects, we were not able to retrieve D$_n$\,4000 measurements due to one or a combination of the following: one of the D$_n$\,4000 bands being partially blanked by bad pixels, emission line contamination from [NeIII], there being no evidence of a break or lack of the reliable spectrograph throughput corrections. 

Various common forbidden lines can also be seen in the spectra of the 70\,$\mu$m sample, originating either in interstellar low-density PhotoDissociation Regions
(PDRs) or the narrow line region (NLR) in the vicinity of an AGN. The luminosity and ionisation state of these lines is related to the strength of ionising UV flux. More specifically, high ionisation states such as [NeV], which we detect in one object, are a consequence of a hard radiation field and hence unambiguous indicators of the presence of an AGN.  The strong [OIII] doublet appears in the 0.3$\lesssim$\,z\,$\lesssim$0.8 spectral range. The 70\,$\mu$m sources display various strengths of the [OIII], [OII], [NeIII] and H\,$\beta$ features and for the purpose of our study we are able to examine the [OIII]/H\,$\beta$ and [NeIII]/[OII] ratios for 33 70\,$\mu$m sources in total. 

In sections \ref{sec:OIII_Hb}, \ref{sec:NeIII_OII} and \ref{sec:Hdelta} we examine the [OIII]/H\,$\beta$ and [NeIII]/[OII] ratios, as well as the behaviour of the H$\delta$ absorption line and its relation to the 4000$\rm \AA$ break and compare with equivalent measurements for the control galaxies matched to each 70\,$\mu$m source. Note, that for all calculations we rely on flux ratios between features at similar wavelength, so the need for extinction correction is eliminated.

\begin{figure}
\epsfig{file=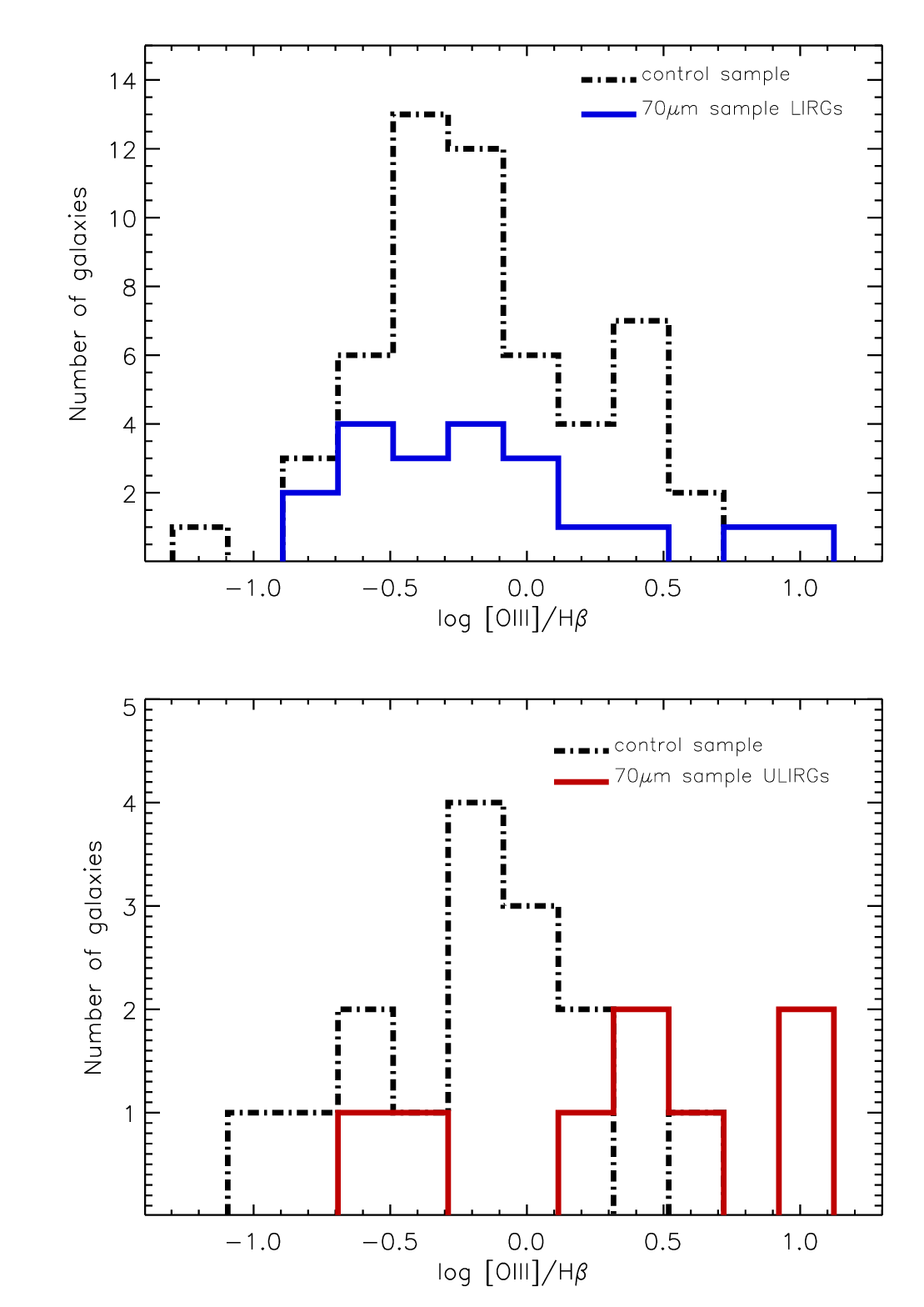,width=8.7cm}
\caption{Distribution of the [OIII]/H$\beta$ ratio for objects in the
 70\,$\mu$m sample (solid lines) --- top panel for the LIRGs and lower panel for the ULIRGs --- and their corresponding control galaxies (black dot-dashed lines).}
\label{fig:Hb_O3}
\end{figure}

\begin{figure}
\epsfig{file=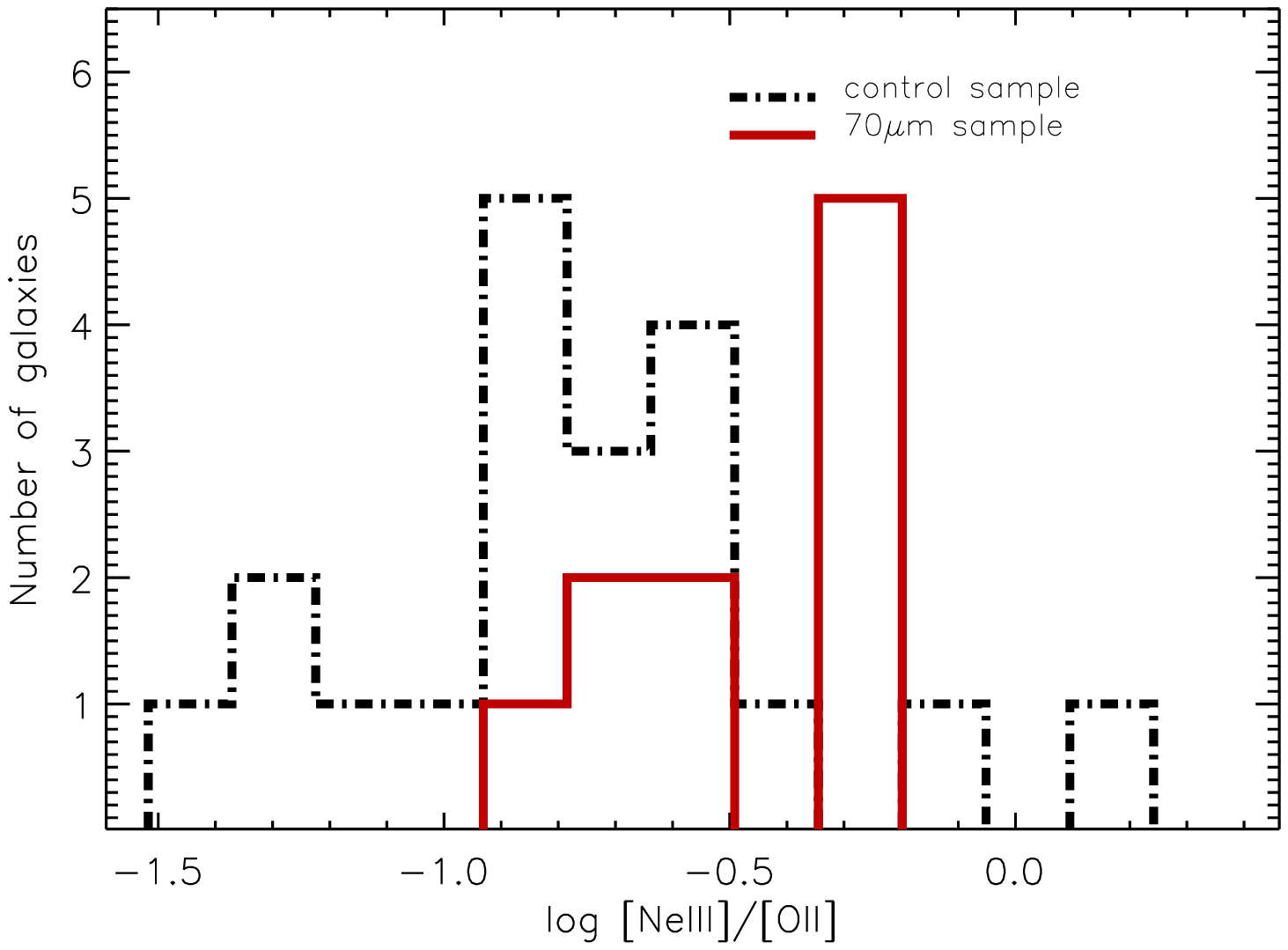,width=8.7cm}
\caption{Distribution of the [NeIII]/[OII] ratio for objects in the
 70\,$\mu$m sample (red solid line) and their respective control galaxies (black dot-dashed line). }
\label{fig:Ne3_O2}
\end{figure}

\subsection{The [OIII]/H\,$\beta$ ratio}
\label{sec:OIII_Hb}

In traditional classification schemes (e.g. Ho, Filippenko $\&$ Sargent
1997\nocite{HFS97}) the [OIII]/H$\beta$ ratio has been used to distinguish starbursts
and Seyferts (e.g. Baldwin, Wampler $\&$ Burbridge 1981\nocite{BWB81}; Veilleux $\&$ Osterbrock 1987\nocite{VO87}),
with higher ratios found in hybrid AGN/Starburst galaxies (e.g. Caputi et
al. 2008\nocite{Caputi08}) and a typical [OIII]/H$\beta$\,$>$\,3 seen in Seyfert
galaxies. Here, we are able to measure [OIII]/H$\beta$ for 20 LIRGs and 8 ULIRGs from the 70\,$\mu$m sample and associated control sources (Fig. \ref{fig:Hb_O3}). For the remaining sources one or both of these lines are out of range. 

The LIRGs appear to have approximately the same log\,[OIII]/H$\beta$ mean value as the control galaxies (Fig. \ref{fig:Hb_O3}, top panel), roughly at -0.25, consistent with the line emission being linked to stellar processes, apart from 2 objects with [OIII]/H$\beta$\,$>$5. Five LIRGs have the H$\beta$ emission line out of range, but the [OIII] marginally in range, however the latter is weak and not measurable in all but one object. In terms of the control sample, a tail-off to high [OIII]/H$\beta$ ([OIII]/H$\beta$\,$>$\,3) is evident, possibly due to Seyfert-types which are likely a few percent of all galaxies at these redshifts. With respect to the ULIRGs (Fig. \ref{fig:Hb_O3}, lower panel), there is more spread in the two distributions, but with the 70\,$\mu$m sample having higher mean [OIII]/H$\beta$. Half the ULIRGs have [OIII]/H$\beta$\,$<$\,3, below the rough dividing line between AGN and star-forming galaxies and half have 3$<$\,[OIII]/H$\beta$\,$<$\,10 typically seen in AGN.
The 5 LIRGs and ULIRGs with [OIII]/H$\beta$\,$>$3 ratios are considered in subsequent work on the AGN fraction (section \ref{sec:agn}).

\subsection{The [NeIII]/[OII] ratio}
\label{sec:NeIII_OII}

The [NeIII]/[OII] ratio is a metallicity indicator, where the negative correlation with metallicity arises due to the atmospheres of more metal-rich stars absorbing a greater fraction of ionising photons, softening the radiation field and decreasing the abundance of the higher ionisation states (e.g. Nagao, Maiolino $\&$ Marconi 2006\nocite{NMM06}). [NeIII]/[OII] is also an ionisation discriminator with AGN showing higher ratios on average, typically $\sim$\,0.4 for type\,2 and $>$1 for type\,1 (e.g. Nagao et al. 2002\nocite{Nagao02}).

The [NeIII]/[OII] ratio for our sample can only be examined for the
z$>$0.7 sources, so all, apart from two, of the 10 sources with reliable measurements are ULIRGs (table \ref{table1} and Fig. \ref{fig:Ne3_O2}). The majority of the control galaxies
are assembled between -1.5$<$log\,[NeIII]/[OII]$<$-0.4, which is the
range one would expect for distant star-forming galaxies (e.g. Perez-Montero et al. 2009\nocite{PM09}), with 2 galaxies at log\,[NeIII]/[OII]$>$-0.5, which are possibly Seyferts or metal-poor. The 70\,$\mu$m sources have -1$<$log\,[NeIII]/[OII]$<$-0.2 in two groups which peak at $\sim$\,-0.7 and $\sim$\,-0.3. The first group of 5 sources is in agreement with the average [NeIII]/[OII] ratio of the control sample. On the other hand, the peak at log\,[NeIII]/[OII]$>$\,-0.4, occupied by the remaining 5 70\,$\mu$m sources, is degenerate as it is consistent with both a type 2 AGN regime (e.g. Nagao et al. 2002\nocite{Nagao02}) and a low metallicity regime (e.g. Perez-Montero et al. 2007\nocite{PM07}).

\begin{figure}
\epsfig{file=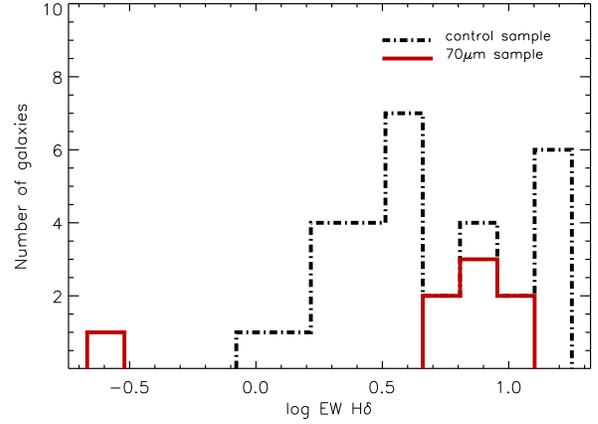,width=8.7cm}
\caption{The distribution in H$\delta$ absorption line EW. Red solid lines: the 7 sources from the 70$\mu$m sample which are in the right redshift range for reliable measurements (see section \ref{sec:spectra}). Black dot-dashed lines: their corresponding control galaxies.}
\label{fig:HD}
\end{figure}

\begin{figure}
\epsfig{file=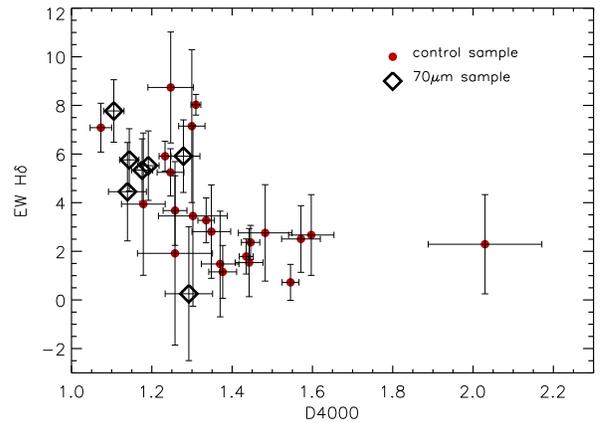,width=8.7cm}
\caption{Plot of H$\delta$ EW vs D$_n$4000. Red filled circles: the control sample, black diamonds: the 70\,$\mu$m sample.}
\label{fig:HD_D4000}
\end{figure}

\subsection{The H$\delta$ absorption line and 4000\,$\rm \AA$ break}
\label{sec:Hdelta}

Figure \ref{fig:HD} compares the H\,$\delta$ absorption line for the 7
sources from the 70\,$\mu$m sample which are in the right redshift
range to have good line measurements, to their respective control
galaxies. The presence of higher order Balmer absorption lines are
indicative of recent star-forming activity (e.g. Dressler $\&$ Gunn
1983), with the peak in H\,$\delta$ absorption occuring at about
$\sim$\,0.5\,Gyr, followed by a decrease during the
subsequent 1-10\,Gyr epoch. The sources in our
sample have higher H\,$\delta$ EWs than the bulk of the control
sample, implying a younger mean stellar population. Nevertheless, H$\delta$
EWs in the 5--10 range are age degenerate, corresponding
to systems that are either very young ($\sim$\,10\,Myr) or
$\sim$1 Gyr old (e.g. Kauffmann et al. 2003\nocite{Kauffmann03a}, hereafter K03). 
To settle this degeneracy we measure the narrow 4000\,$\rm \AA$ break (D$_n$\,4000) break and examine its strength against the H\,$\delta$ absorption
EW in figure \ref{fig:HD_D4000}. The 4000\,$\rm \AA$ break is a monotonic function of age, small for young stellar populations and large for old, metal-rich galaxies.  We find that both the control and 70\,$\mu$m sample lie where expected
for star-forming galaxies with EW(H\,$\delta$)\,$>$\,1 and
D4000\,$>$\,1, however their spread over the D$_n$4000-H$\delta$ plane
is substantially different: the control sample spans a large
part of the plane, whereas the 70\,$\mu$m sources have low
D$_n$4000 ($<$\,1.3) and most have high EW (H\,$\delta$)  ($>$\,5) indicative of young stellar populations and consistent with a starburst mode of star formation (e.g. K03). Similar values of D$_n$4000 and H\,$\delta$ are also
reported by Marcillac et al. (2006\nocite{Marcillac06}) and Caputi et al. (2008\nocite{Caputi08})) who
examine distant IR-selected galaxies. Caputi et al. (2008) find them
to have the smallest D$_n$4000 values amongst a large sample of
sources from the COSMOlogical evolution Survey (COSMOS), on par with
the 70\,$\mu$m sample occupying the lower end of the D$_n$4000
distribution of DEEP\,2 control galaxies.

\begin{figure*}
\epsfig{file=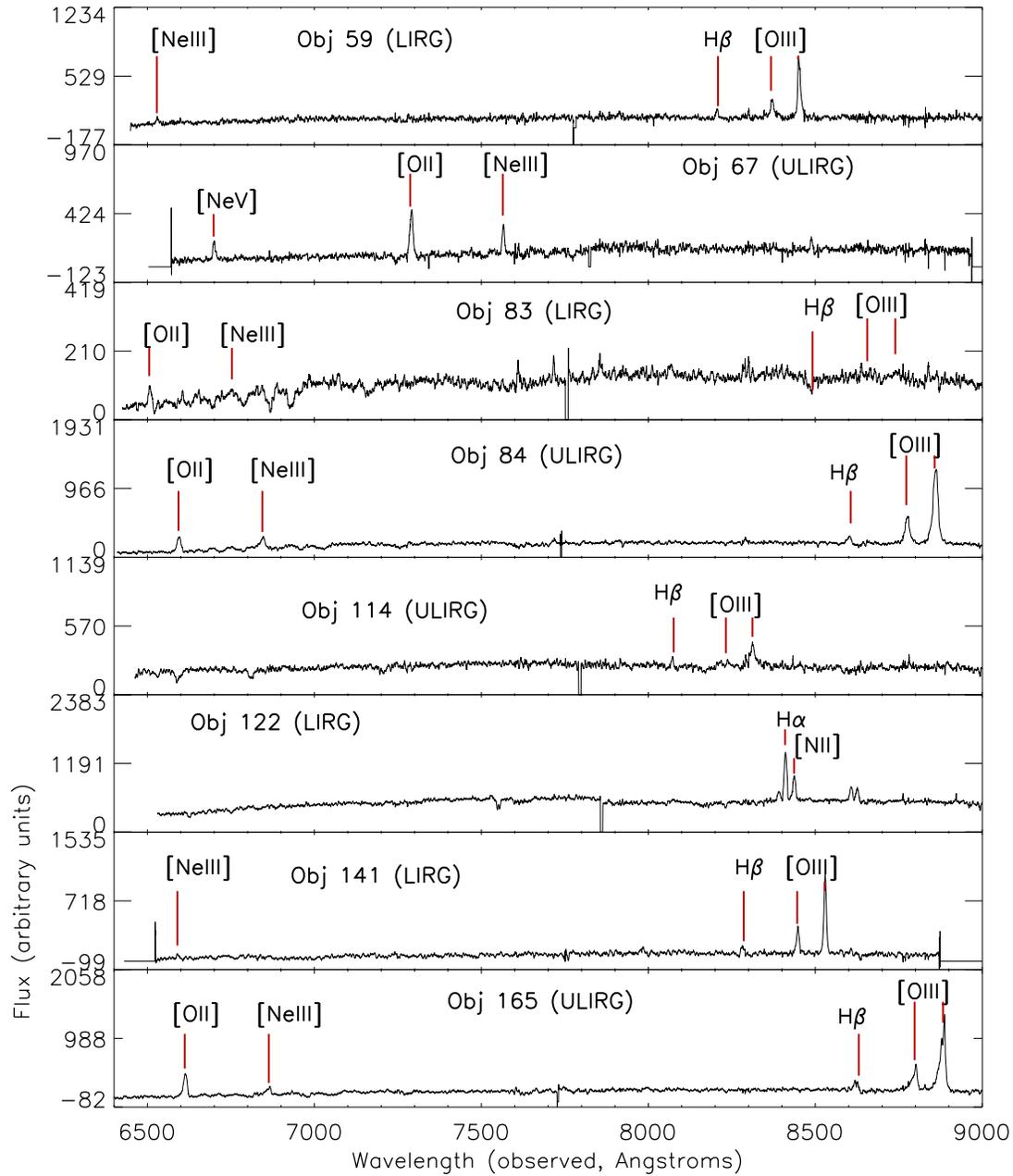,width=15cm}
\caption{DEEP\,2 spectra for the 8 AGN candidates, whose SEDs are shown in figure \ref{fig:xrayIR_SEDs}.} 
\label{fig:spectra}
\end{figure*}

\begin{figure*}
\epsfig{file=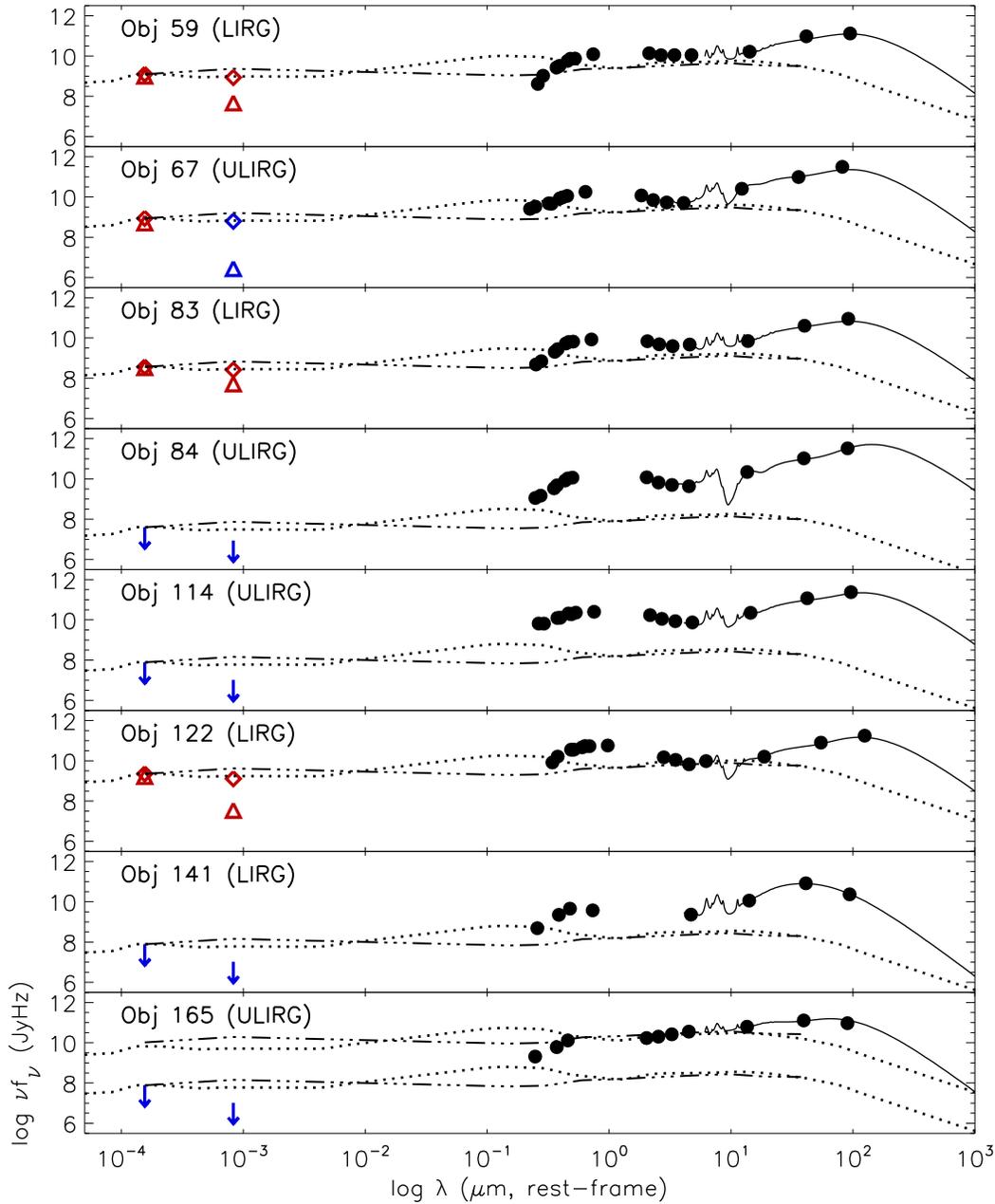,width=15cm}
\caption{The X-ray to IR SEDs for the 8 objects in the sample which show at least one AGN signature; black filled circles: optical/IR data,  black
 solid line: SK07 IR model SED templates (normalised to 24\,$\mu$m); dotted
 line: radio quiet QSO SED from Elvis et al. (1994) (normalised to the absorption-corrected
 hard-band X-ray flux), dash-dot line: SED of AGN NGC\,7213 from Kuraszkiewicz et al. (2001) (normalised to the absorption-corrected
 hard-band X-ray flux), diamonds: hard and soft fluxes corrected for absorption, triangles: hard and soft fluxes not corrected for
 absorption (red for measured flux, blue for upper limit). When there is no X-ray detection the AGN SEDs are normalised to the hard X-ray upper limits. For object 165, which does not have an X-ray detection, but has a power-law near-IR continuum, the templates are normalised to both the hard X-ray upper limit and the near-IR continuum.} 
\label{fig:xrayIR_SEDs}
\end{figure*}

\begin{figure*}
\epsfig{file=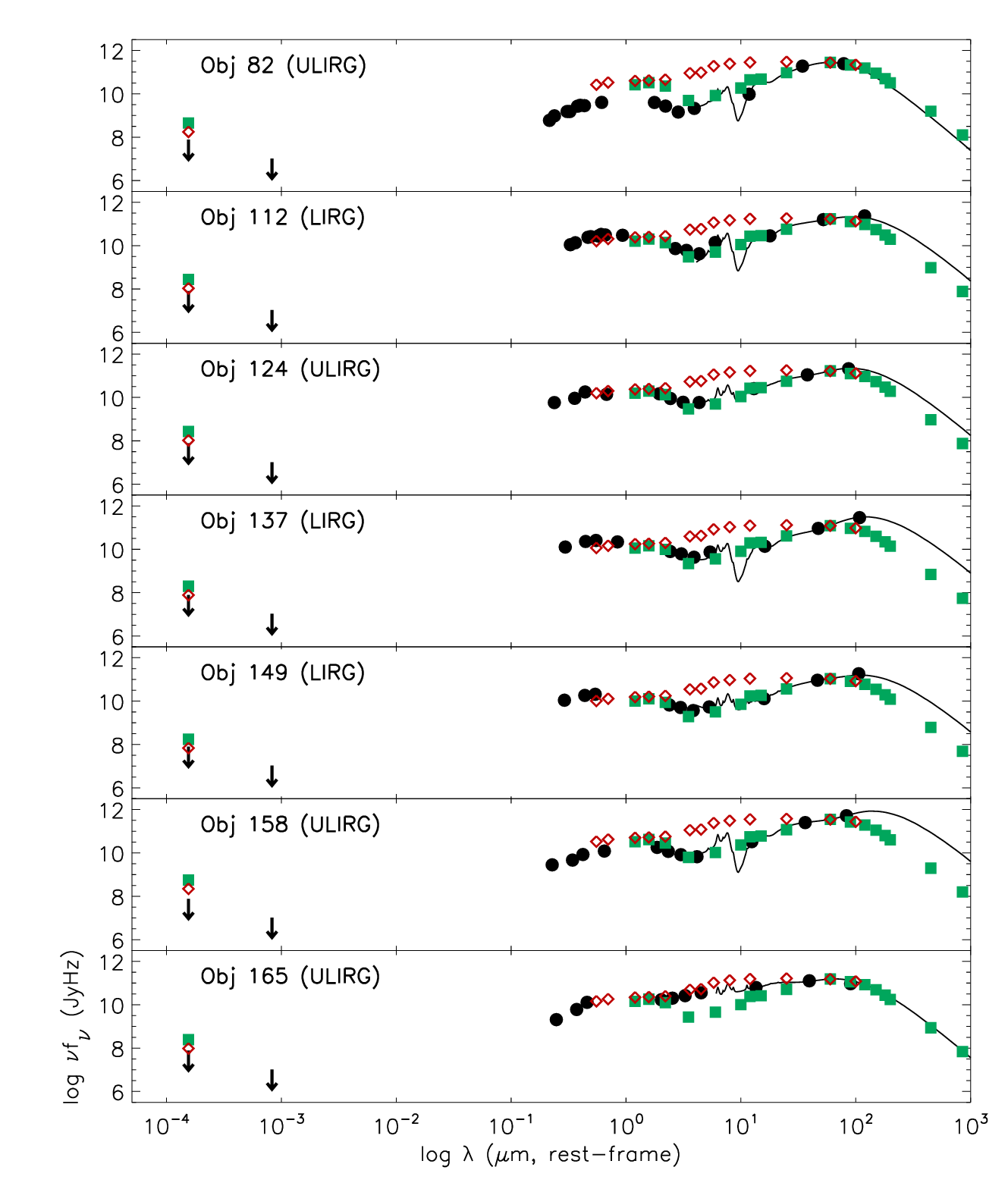,width=15cm}
\caption{The X-ray to IR SEDs of 6 objects with no AGN signatures (the LIRGs and ULIRGs in
Fig. \ref{fig:spectra_all}) and AGN-candidate object 165; black upper limits: X-ray data; black filled circles: optical/IR data;  black solid line: SK07 IR model SED templates (normalised to 24\,$\mu$m). The red open diamonds represent the panchromatic SED of NGC\,1068 --- photometry for the SED from Sandage (1973); Spinoglio et al. (1995); Stickel et al. (2004); Cappi et al. (2006) and Howell et al. (2007). The green filled squares represent the panchromatic SED of NGC\,6240 --- photometry for the SED from Allen (1976); Soifer et al. (1989); Iwasawa $\&$ Comastri (1998); Spinoglio et al. (1995); Klaas et al. (2001) and Lutz et
al. (2004). The SEDs of NGC\,1068 and NGC\,6240 are normalised to the SK07 template at the IRAS 60$\mu$m photometry, in order to be consistent with our original sample
selection at 70$\mu$m. Note that for both local galaxies including the soft X-ray flux is not relevant, as it is dominated by emission lines from photoionized plasma, with the underlying X-ray continuum completely absorbed. The hard X-ray flux that we plot here is observed and not corrected for absorption.}
\label{fig:xrayIR_SEDs_nonAGN}
\end{figure*}

\nocite{Sandage73}
\nocite{Spinoglio95}
\nocite{Stickel04}
\nocite{Cappi06}
\nocite{Howell07}
\nocite{Allen76}
\nocite{Soifer89}
\nocite{Iwasawa_Comastri98}
\nocite{Spinoglio95}
\nocite{Klaas01}
\nocite{Lutz04}

\section{The AGN content of the 70\,$\mu$\MakeLowercase{m} population}
\label{sec:agn}

\subsection{The AGN fraction}
\label{sec:fraction}

In order to calculate the AGN fraction, we use four diagnostics --- hard X-ray emission, high ($>$3) [OIII]/H$\beta$ ratio, the presence of [NeV] lines and a power-law near/mid-IR continuum --- with the requirement that at least one is satisfied for an object to be classed as an AGN candidate; see table \ref{table:AGN} and figures \ref{fig:spectra} and \ref{fig:xrayIR_SEDs}.

The X-ray fraction in the 70\,$\mu$m sample amounts to 16\,($\pm$6) per cent: two objects are
detected in the full X-ray band only (objects 40 and 84), whereas 8
have detections in at least one other band (objects 43, 56, 59, 67,
83, 93, 122 and 147). However, as star formation can also be a source
of X-ray emission (see section \ref{sec:IR}) we only consider sources with hard X-ray detections to be
AGN candidates --- objects 59, 67, 83 and 122 --- especially since for
the remaining sources X-ray emission is consistent with the
star-forming relations in figure \ref{fig:xrayIR}. These 4 sources also have log\,L$_{2-10\rm keV}$\,$>$\,42, which, as previously mentioned, usually indicates that hard X-ray emission is AGN-dominated. 

A power-law type near/mid-IR (3\,$\lesssim$\,$\lambda$\,$\lesssim$\,20\,$\mu$m) continuum and hence red colours in that part of the SED, are a consequence of emission
from dust heated to near sublimation temperatures. As the strength of
the stellar radiation field is not sufficient to cause this, the only
stellar-related mechanism that could be responsible for a near/mid-IR
excess is dust stochastically heated by shocks from outflows and
supernovae (e.g. Ho et al. 1989\nocite{H89}; Davies, Burston $\&$ Ward
2002\nocite{DBW02}). However, studies have shown that the dust grains responsible
are of very small size and hence lifetimes of the order of only a few
kyr (Dwek 1986\nocite{D86}), making this an unlikely explanation for a
near-IR excess in a broad-band SED. A more likely scenario is that of AGN dust heating and hence direct emission from the torus (Neugebauer 1979\nocite{N79}; Elvis et
al. 1994\nocite{E94}; Lutz et al. 1998\nocite{Lutz98}; Sturm et
al. 2000\nocite{Sturm00}; Klaas et al. 2001\nocite{Klaas01}; Alonso-Herrero et al. 2006\nocite{AH06a}). 
There is one object in the 70\,$\mu$m sample with a power-law
near/mid-IR continuum and red colours in all IRAC bands (3.6--8\,$\mu$m) --- object 165 (see SED in
Fig. \ref{fig:xrayIR_SEDs}). It also has a high ($>$3)
[OIII]/H$\beta$ ratio; see table \ref{table:AGN}. The remaining AGN candidate sources have starburst-type SEDs, although this does not exclude an AGN contribution to the near/mid-IR. In terms of AGN spectroscopic signatures, object 67 has an evident
[NeV] line and objects 84, 114, 141, 122, 165 have high
[OIII]/H$\beta$ ratios (objects 67 and 122 are also hard X-ray
emitters). Object 83, one of the hard X-ray emitters, has a spectrum that is almost devoid of emission lines, which implies complete
obscuration or absence of the narrow line region and could be an
example of a host-obscured source (e.g. Martinez-Sansigre
2006\nocite{MS06}), or a source with near 4$\pi$ torus covering factor
(Imanishi et al. 2007\nocite{Imanishi07}). 

In total, 8 out of 61 sources in the 70\,$\mu$m sample show evidence for an active nucleus by satisfying at least one diagnostic, making the AGN fraction 13 per cent, or 11 and 23 per cent when the LIRGs and
ULIRGs are considered seperately. Note that this is the fraction of
sources which host an AGN, not the fraction of sources powered by AGN and it is likely a lower
limit, as it is possible that a number of AGN without spectroscopic or photometric signatures could be missed --- see section \ref{sec:compton_thick} for a
discussion.

\begin{table*}
\centering
\caption{The 8 objects identified to host AGN. Columns 1--6: object ID, infrared luminosity class,
  SED type, hard X-ray luminosity, the [OIII]/H$\beta$ ratio, a
  visible NeV line. L$_{IR}$/L$_{AGN}$ is the ratio of the integrated energy in the
  8-1000\,$\mu$m spectral region to the integrated energy in the QSO
  template from Elvis et al. (1994) when normalised to the absorption-corrected hard X-rays or X-ray upper limit (column 7) and the IRAC 8\,$\mu$m photometry (column 8). The latter represents the maximum AGN contribution as it is constrained by our infrared photometry. Column 9 identifies the primary energy source of the galaxy. Given the power-law-type SED of object 165, the AGN likely has a substantial contribution to the galaxy's energy budget, however the L$_{IR}$/L$_{AGN}$ ratio shows that the starburst component is still the primary energy source.}
\begin{tabular}{|c|c|c|c|c|c|c|c|c|}
ID & L$_{IR}$ class & SED type & log L$_{2-10 \rm keV}$(erg/s)&[OIII]/H$\beta$ & [NeV] & L$_{IR}$/L$_{AGN }$ (X) &  L$_{IR}$/L$_{AGN}$ (8) & Powered by\\
\hline
59 & LIRG & starburst & 42.9 & 7& out of range&23&10&starburst\\
67 & ULIRG & starburst & 43.3 & out of range& yes&59&40&starburst\\
83 & LIRG & starburst & 42.6 & too weak & out of range&42&13&starburst\\
84 & ULIRG & starburst & not detected & 9 & out of range&2448&90&starburst\\
114 & ULIRG & starburst & not detected & 3 & out of range&640&27&starburst\\
122 & LIRG & starburst & 43.4 &out of range& out of range &14&13&starburst\\
141 & LIRG & starburst & not detected & 6 & out of range&192&26&starburst\\
165 & ULIRG & power-law & not detected & 8&out of range& 550&6&starburst/AGN\\
\hline 
\end{tabular}
\label{table:AGN}
\end{table*}

\subsection{The AGN contribution to the energy budget}
\label{sec:energy_budget} 

For the 8 sources which are found to host an active nucleus, we investigate the AGN contribution to the total energy budget with the aid of SED templates; figure
\ref{fig:xrayIR_SEDs} shows the sources' panchromatic SEDs with X-ray fluxes in the hard and
soft bands, B, R, I fluxes from DEEP\,2, J and K fluxes where available, IRAC and MIPS data and the SK07 IR template best matched to each object (see section \ref{sec:previous_work}). We use \emph{Webpimms}, a NASA High Energy Astrophysics Science Archive Research
Center (HEASARC) tool, to estimate absorbing column densities (N$_H$) and correct the fluxes for absorption, via our estimated hardness ratios and a $\Gamma$=1.9 photon index. These are shown in table \ref{table1} where the upper and lower values for the column density are calculated using the upper and lower values of the hardness ratios. We use the average SED for a radio quiet QSO from Elvis et
al. (1994\nocite{E94}) and that of a lower luminosity, lower Eddington-ratio AGN, NGC\,7213 from Kuraszkiewicz et al. (2003\nocite{Kuraszkiewicz03}) which lacks the `big blue bump'. Assuming that hard X-rays are a measure of AGN power, both templates are 
normalised to the absorption corrected hard X-ray flux or the hard-band upper limit if the object either has no
X-ray detection or it is detected only in the full-band. In order to separate the AGN and
host galaxy SED components we use templates which have little or no contribution from the host galaxy and therefore represent integrated
energy from the AGN. Note that because these templates represent unabsorbed AGN, they would not be expected to match our observed spectrum in the optical/UV according to the absorbing columns we calculate. Nevertheless and assuming the unified scheme holds (Urry $\&$ Padovani
1995\nocite{UP95}), the difference between an unabsorbed and an absorbed AGN
SED should be far more obvious in the soft X-ray/EUV/UV and optical,
whereas the contribution from the torus in the mid and far-IR, where it is optically thin, should
vary much less with orientation. Consequenctly, since our aim is to calculate the integrated AGN energy by normalising the templates to our sources' photometry, it is not appropriate to use absorbed AGN templates where a large part of the spectrum would be partially or completely wiped out. This is particularly important for higher luminosity AGN, where most energy comes out in the UV and optical. Moreover, unabsorbed templates allow us to evaluate the strength of the stellar bump in comparison to the AGN in the 1-2\,$\mu$m region. 

The SEDs of all 8 sources (Fig. \ref{fig:xrayIR_SEDs}) are characterised by a far-IR bump significantly more luminous than the AGN templates in that
part of the spectrum. For objects 84, 114 and 141 the entire optical/IR SED is displaced from the AGN templates and is 2-4 orders of
magnitude more luminous, an indication that the AGN is of low
luminosity and therefore unlikely to contribute substantially to the
galaxies' energy budget. For the hard X-ray emitters, objects 59, 67, 83
and 122, the AGN templates agree better with the mid-IR SEDs, indicating that some part of the galaxies' near/mid-IR luminosity must be coming from the AGN. Object 165 is not detected in the X-rays, but it is classed as hosting an AGN from spectral signatures and mid-IR continuum slope. Accordingly it is more appropriate to
normalise the AGN template to the near/mid IR continuum, as the power-law SED slope in that region is characteristically emission from the torus, showing that even if the AGN is not detected in the X-rays, it could be contributing substantially to the source's energy budget.

We examine the possibility that the far-IR bump results from the high frequency light from the AGN being absorbed and
re-radiated into the IR. Assuming that X-ray emission scales with AGN power, then to power a LIRG or ULIRG
(L$_{IR}$\,=\,10$^{11}$-10$^{13}$\,L$_{\odot}$) and hence dominate
emission in the far-IR, we would expect an AGN luminosity of at
least a few times 10$^{44}$\,erg/s and assuming that a few per cent is emerging in the X-rays (e.g. Elvis et al. 1994), an X-ray luminosity of at least 10$^{43}$\,erg/s for the low luminosity IR sources and at least 10$^{44}$\,erg/s for the more luminous LIRGs and ULIRGs. None of our objects satisfy these criteria. In fact, quite simply integrating under the SEDs, reveals that in all cases the IR bump (8--1000\,$\mu$m) has between 1--3 orders of magnitude
more energy than the AGN/QSO templates (defined as the $L_{IR}$/$L_{AGN}$ ratio). It is possible that the column densities, and hence absorption corrections, derived from our hardness ratios could be underestimated if there is a contribution from scattered radiation in the soft X-ray band for heavily absorbed sources. In order to address this, we estimate a maximum value for $L_{IR}$/$L_{AGN}$ by normalising the AGN template to our 8\,$\mu$m IRAC photometry. We find that the IR SED overtakes emission from the AGN by factors of 6--90 and hence conclude that a) with the exception of object 165, where $L_{IR}$/$L_{AGN}$ is less than an order of magnitude, for the remaining objects the AGN contributes less than 10 per cent to the IR energy budget and b) all sources are primarily powered by star-formation --- see table \ref{table:AGN}.

\begin{table*}
\begin{minipage}{126mm}
\centering
\caption{Table of stacked fluxes (erg/s/cm$^2$) and hardness ratios for the 70\,$\mu$m sample minus the 8 AGN candidates (see section  \ref{sec:fraction}). They are split into 3 luminosity classes (starbursts, LIRGs and ULIRGs; see section \ref{sec:previous_work} for details on the sources' infrared properties). The second column indicates how many objects were
 used in the stacking --- some were removed because they were too close to a source which could have contaminated the signal or they were associated with an X-ray source. The SB group does not have a formal detection in the hard band, so we quote the 3$\sigma$ upper limit. The SB hardness ratio is derived using the counts in the soft band and 3$\sigma$ upper limit counts in the hard band. }
\begin{tabular}{|c|c|c|c|c|}
Class & Objects used & f$_{0.5-2}$ & f$_{2-10}$ & HR\\
\hline 
Starbursts & 8 out of 9 & 8.77E-17 (8$\sigma$) & 2.35E-16 (3$\sigma$ upper limit)&-0.3 (upper limit)\\
LIRGs & 26 out of 31 & 5.93E-17 (10$\sigma$)& 1.6e-16 (4$\sigma$)& -0.3\\
ULIRGs & 9 out of 13&9.82E-17 (10$\sigma$)&2.33E-16 (3$\sigma$)& -0.4\\
\end{tabular}
\label{stacking_nonagn}
\end{minipage}
\end{table*}

\subsection{Obscured AGN}
\label{sec:compton_thick}

Our results indicate that low luminosity (L$_{2-10keV}$\,$\sim$\,10$^{42}$\,erg/s) AGN with moderate obscuration (N$_H$\,$\sim$\,10$^{23}$\,cm$^{-2}$) can be detected in our survey up to z$\sim$0.5.  Active nuclei obscured by higher column densities are potentially missed even if more luminous, implying that the AGN fraction that we calculate is a lower limit. Nevertheless, the hardness ratios we calculate for our stacked sources (table \ref{stacking}) are soft (HR$<$-0.2) indicating a minor (if any) contribution from obscured AGN to the X-ray signal, making our AGN fraction relatively robust --- this is further confirmed by the stacking signal from the same sample of sources minus the 8 AGN candidates (see table \ref{stacking_nonagn}). In addition, even without X-ray detections, it is unlikely that we would mis-classify AGN-powered 70\,$\mu$m sources as starburst-dominated. As outlined in sections \ref{sec:fraction} and \ref{sec:energy_budget}, AGN and starburst SEDs are highly divergent in the near/mid IR and therefore if one of our L$>$10$^{11}$\,L$_{\odot}$ sources were to host a L$_{2-10keV}$\,$>$\,10$^{44}$\,erg/s, haevily obscured AGN which could potentially power the galaxy but is not detected in the X-rays, its SED would show a strong near/mid-IR continuum: hot dust emission from the torus. 

To demonstrate this further we compare the
SEDs of 6 objects with no AGN signatures (the LIRGs and ULIRGs in
Fig. \ref{fig:spectra_all}) and AGN-candidate object 165 to the SEDs
of the local IR galaxies NGC\,6240 and NGC\,1068 known to host compton thick AGN
(Fig. \ref{fig:xrayIR_SEDs_nonAGN}). The diverse nature of NGC\,6240 and NGC\,1068 is unambiguously mirrored in their panchromatic SEDs, which are in agreement over the optical and far-IR parts but diverge significantly in the near/mid-IR. 
NGC\,6240 is a SB/AGN hybrid with a binary compton thick ($\sim$10$^{24}$
cm$^{-2}$), $\sim$10$^{44}$ erg/s QSO, but the
galaxy is nevertheless primarily powered by a starburst (e.g. Vignati et al. 1999\nocite{Vignati99}; Komossa $\&$ Schultz 1999\nocite{Komossa_Schulz99}; Klaas et al. 2001\nocite{Klaas01}), whereas NGC\,1068 is a type-II Seyfert (L$_{2-10keV}$\,$>$10$^{44}$ erg/s) with an additional far-IR starburst component (e.g. Le Floc'h 2001\nocite{LeFloch01}; Spinoglio et al. 2005).
We normalise their SEDs to the SK07 template at the IRAS 60$\mu$m photometry, in order to be consistent with our original sample
selection at 70$\mu$m. We do not consider the soft X-ray flux, which has limited diagnostic value for such heavily obscured AGN embedded in star-forming galaxies. The hard X-ray flux we show in the SEDs of NGC6240 and NGC1068 is observed and hence not corrected for absorption.

We see from Fig. \ref{fig:xrayIR_SEDs_nonAGN} that if there are any sources in the 70\,$\mu$m sample with IR to X-ray (L$_{60 \mu m}$/L$_{2-10keV}$) ratios similar to NGC\,6240 and NGC\,1068 then they would on average be detected in our X-ray survey. However, even if such compton thick AGN are missed, the near/mid-IR part of the SED would show unambiguous signatures of their presence if they were to be significant contributors to the galaxy's energy budget. Not surpsisingly, apart from object 165, whose SED is better represented by that of NGC\,1068, the remaining 70\,$\mu$m sources are in agreement with the starburst-type SED of NGC\,6240.

\onecolumn
\begin{landscape}
\begin{center}
\begin{longtable} {|c|c|c|c|c|c|c|c|c|c|c|c|c|c|c|}
\caption{The 70\,$\mu$m sample (61 objects): IDs from the 70\,$\mu$m EGS survey and the DEEP\,2 survey, redshift, R-band magnitude (AB), total infrared luminosity, X-ray flux in the soft and hard bands ($\times$10$^{-16}$\,erg\,s$^{-1}$\,cm$^2$), hardness ratios, column density (cm$^{-2}$) with the upper and lower limits, line ([OII], [NeIII] H$\beta$, [OIII]) fluxes (log W/m$^2$), H$\delta$ absorption line equivalent width ($\rm \AA$) and the reduced 4000\,$\rm \AA$ break. `(l)' indicates a 3$\sigma$ upper limit for the fluxes. Column densities were calculated with $\Gamma$=1.9. `(l)' in the N$_H$ column indicates that with a $\Gamma$=1.9 the estimated column density is lower than the Galactic N$_H$ of $\sim$\,2\,$\times$\,10$^{20}$. The upper and lower limits of N$_H$ are calculated from the upper and lower values of the hardness ratio.} \\
\hline
ID & DEEP ID & z & R mag & $L_{IR} \times 10^{10}(L_{\odot})$ &f$_{0.5-2}$ & f$_{2-10}$ & HR & N$_H$&f$_{[OII]}$&f$_{[NeIII]}$&f$_{H\beta}$&f$_{[OIII]}$&EW$_{H\delta}$&D$_n$4000\\
\hline \hline 
\endfirsthead
\multicolumn{15}{|c|}{{\bfseries \tablename\ \thetable{} -- continued from previous page}} \\
\hline
ID & DEEP ID & z & R mag & $L_{IR} \times 10^{10}(L_{\odot})$ & f$_{0.5-2}$ & f$_{2-10}$ & HR & N$_H$&f$_{[OII]}$&f$_{[NeIII]}$&f$_{H\beta}$&f$_{[OIII]}$&EW$_{H\delta}$&D4000\\
\hline \hline 
\endhead
\hline \multicolumn{15}{|r|}{{Continued on next page}} \\ 
\hline
\endfoot
\hline \multicolumn{15}{|r|}{{End of table}} \\ 
 \hline
\endlastfoot
\hline \multicolumn{13}{|c|}{{\bf{STARBURSTS}}} &\\
\hline
      20&    14017095&0.33&19.70&8.72$\pm$0.11&-&-&-&-&- &- &- &- &- &-\\
      46&    13056925&0.22&19.72&9.24$\pm$0.04&-&-&-&-&- &- &- &- &- &-\\
      51&    13058203&0.22&18.59&4.37$\pm$0.23&-&-&-&-&- &- &- &- &- &-\\
      56&    13041622&0.20&18.70&9.84$\pm$0.49&9.04$\pm^{1.7}_{1.55}$&13.7(l)&-0.48$\pm^{0.17}_{0.2}$ &(l) \,$^{(\rm l)}_{(\rm l)}$&-&-&-&-&-&-\\
      81&    13009920&0.23&19.65&5.38$\pm$0.38&-&-&-&-&- &- &- &-&-&-\\
     118&    12008271&0.24&20.30&3.77$\pm$0.30&-&-&-&-&- &- &- &-&-&-\\
     125&    11051641&0.25&20.72&8.28$\pm$0.12&-&-&-&-&- &- &- &-&-&-\\
     144&    11033888&0.19&19.35&3.62$\pm$0.18&-&-&-&-&- &- &- &-&-&-\\
     155&    11020745&0.17&18.60&2.74$\pm$0.30& -&-&-&-&- &- &- &-&-&-\\
\hline \hline
\hline \multicolumn{13}{|c|}{{\bf{LIRGs}}} &\\
\hline
      33&    14012882&0.34&19.80&18.72$\pm$2.87&-&-&-&-&-&-&-& -19.23&-&-\\
      37&    13063596&0.36&19.49&45.23$\pm$0.46&-&-&-&-&-&-& -18.86&  -19.43&-&-\\
      40&    13063597&0.30&18.36&41.96$\pm$2.10&1.75(l)&10.4(l)&-0.24$\pm^{0.3}_{0.3}$&5.4$\times10^{21}$\,$^{1.6 \times 10^ {22}}_{(\rm l)}$&-&-&-&-19.70&-&-\\
      41&    14007624&0.45&20.88&80.59$\pm$7.03&-&-&-&-&-&  -&  -19.61&  -20.06&-&-\\
      43&    13063920&0.56&20.73&32.42$\pm$3.09& 1.13$\pm^{0.97}_{0.63}$&2.23(l)&-0.66$\pm^{0.33}_{0.08}$&(l)\,$^{4 \times 10^ {21}}_{(\rm l)}$&-&  -&  -19.03&  -19.39&-&-\\
      55&    13049741&0.67&21.56&84.70$\pm$5.68&-&-&-&-&    -&  -&  -19.33&  -20.15&  5.52& 1.19\\
      59&    13050479&0.69&22.33&82.16$\pm$2.77&4.99$\pm^{1.49}_{1.32}$ &106$\pm^{12.4}_{11.7}$&0.61$\pm^{0.08}_{0.1}$ &$1.1\times10^{23}$\,$^{1.15 \times 10^ {23}}_{8 \times 10^ {22}}$&-&  -19.86&  -19.70&  -18.88&0.25&1.29\\
      68&    13034644&0.35&19.65&33.96$\pm$1.50&-&-&-&-&-&-&-&-19.12&-&-\\
      72&    13035302&0.39&20.44&13.77$\pm$1.18&-&-&-&-&-&-&-& -20.07&-&-\\
      76&    13017944&0.73&22.62&30.96$\pm$2.16&-&-&-&-&-&  -&  -19.54&  -19.68&4.45&1.14\\
      78&    13026857&0.37&20.27&35.92$\pm$0.15&-&-&-&-&-&  -&  -18.91&  -19.58&-&-\\
      83&    13019240&0.75&22.55&43.08$\pm$11.08&5.53$\pm^{1.5}_{1.31}$&35.5$\pm^{8.78}_{7.99}$&0.15$\pm^{0.14}_{0.18}$ &$4.2\times10^{22}$\,$^{5.7 \times 10^ {22}}_{2.6 \times 10^ {22}}$&-19.99&  -20.73& -&  -19.74&-&-\\
      85&    13003379&0.43&19.92&25.80$\pm$0.92&-&-&-&-&-&  -&  -18.82&  -19.54&-&-\\
      94&    12027947&0.55&21.69&76.92$\pm$3.03&-&-&-&-& -& -&  -19.17&  -19.77&-&-\\
      95&    12027969&0.73&22.01&90.08$\pm$29.67&-&-&-&-&-&-&-&-&-&-\\
      98&    12023868&0.29&20.10&10.82$\pm$1.19&-&-&-&-&-&-&-&-&-&-\\
     100&    12024073&0.30&19.33&23.56$\pm$1.57&-&-&-&-&-&-&-&-&-&-\\
     103&    12029058&0.25&19.02&10.90$\pm$0.15&-&-&-&-&-&-&-&-&-&-\\
     110&    12015978&0.42&21.29&33.28$\pm$2.03&-&-&-&-& -&  -&  -20.42&  -20.42&-&-\\
     112&    12020772&0.34&19.83&27.63$\pm$7.78&-&-&-&-&-&  -&  -&  -19.46&-&-\\
     115&    12008273&0.38&20.08&25.12$\pm$2.17&-&-&-&-&-&  -&  -18.77&  -19.24&-&-\\
     120&    12013167&0.57&21.61&30.13$\pm$3.76&-&-&-&-& -&  -&  -19.38&  -19.12&-&-\\
     121&    12008914&0.78&22.07&57.08$\pm$8.60&-&-&-&-&-19.29&  -19.64&  -19.17&  -19.49&  7.77& 1.1\\
     122&    12004450&0.28&19.42&11.80$\pm$2.70&3.56$\pm^{1.51}_{1.28}$&172$\pm^{17.9}_{16.9}$&0.78$\pm^{0.05}_{0.07}$&$7\times10^{22}$\,$^{8 \times 10^ {22}}_{6 \times 10^{22}}$&-&-&-&-&-&-\\
     126&    11051657&0.42&21.16&27.97$\pm$2.23&-&-&-&-&-&-&-&-&-&-\\
     135&    11046414&0.48&20.92&57.20$\pm$0.28&-&-&-&-&-&-&-&-&-&-\\
     137&    11046419&0.48&20.07&67.64$\pm$0.39&-&-&-&-&-&-&-18.61&-18.75&-&-\\
     138&    11039469&0.42&21.13&12.54$\pm$6.27&-&-&-&-&-&-&-19.28&-19.9&-&-\\
     140&    11047086&0.46&19.42&27.08$\pm$5.62&-&-&-&-&-&-&-&-&-&-\\
     141&    11101177&0.70&22.74&36.76$\pm$1.84&-&-&-&-&-&-20.09&-19.58&-18.79&10.14&-\\
     146&    11033882&0.46&20.05&19.64$\pm$2.18&-&-&-&-&-&-&-&-&-&-\\
     147&    11026910&0.67&22.31&85.49$\pm$14.92&1.53$\pm^{0.88}_{0.64}$&2.58(l)&-0.5$\pm^{0.39}_{0.2}$&(l) \,$^{1.8 \times 10^ {22}}_{(\rm l)}$&-&-&-19.74&-20.61&-&-\\
     149&    11027507&0.51&20.34&46.83$\pm$7.07&-&-&-&-&-&-&-18.52&-18.45&-&-\\
     150&    11034350&0.74&21.97&48.99$\pm$8.80&-&-&-&-&-&-&-19.44&-19.82&5.34&1.17\\
     152&    11027541&0.37&19.98&54.70$\pm$0.37&-&-&-&-&-&-&-19.19&-19.42&-&-\\
\hline \hline
\hline \multicolumn{13}{|c|}{{\bf{ULIRGs}}} &\\
\hline
      24&    14011285&0.76&22.07&153.57$\pm$23.59&-&-&-&-&-19.82&-&-20.28&-19.89&-&-\\
      26&    14017213&1.29&23.68&796.20$\pm$51.18&-&-&-&-&-19.38&-20.23&-&-&-&-\\
      58&    13041891&0.69&22.29&196.78$\pm$42.96&-&-&-&-&-&-&-19.67&-20.35&5.91&1.28\\
      67&    13034619&0.96&22.11&235.69$\pm$2.43&0.3(l)&55.2$\pm^{11.9}_{10.4}$&0.93$\pm^{0.01}_{0.07}$&$3.7\times10^{23}$\,$^{3.9 \times 10^ {23}}_{2.8 \times 10^ {23}}$&-19.17&-19.56&-&-&-&-\\
      70&    13026142&0.78&21.30&223.68$\pm$56.86&-&-&-&-&-19.62&-20.32&-&-&-&-\\
      77&    13036021&0.42&19.16&101.04$\pm$0.60&-&-&-&-&-&-&-18.96&-19.52&-&-\\
      82&    13027191&1.04&23.36&275.11$\pm$44.43&-&-&-&-&-19.34&-19.73&-&-&-&-\\
      84&    13019950&0.77&21.97&307.45$\pm$64.62&0.92(l)&4.2(l) &0.32$\pm^{0.27}_{0.4}$&$6.3\times10^{22}$\,$^{1.1 \times 10^ {23}}_{2.3 \times 10^ {22}}$&-19.20&-19.50&-19.34&-18.39&-&-\\
      88&    13019982&0.78&21.51&254.05$\pm$10.51&-&-&-&-&-19.16&-19.51&-19.78&-19.40&-&-\\
      93&    13004291&1.20&22.57&488.97$\pm$74.57&1.59$\pm^{0.92}_{0.67}$&3.82(l)&-0.34$\pm^{0.35}_{0.26}$&$1\times10^{22}$\,$^{5 \times 10^ {22}}_{(\rm l)}$&-19.54&-20.50&-&-&-&-\\
     101&    12028577&0.82&23.63&100.84$\pm$12.61&-&-&-&-&-&-&-&-&-&-\\
     114&    12007915&0.66&20.83&128.59$\pm$2.51&-&-&-&-&-&-&-19.66&-19.18&-&-\\
     124&    11051625&0.85&21.31&173.78$\pm$43.18&-&-&-&-&-19.2&-&-&-&-&-\\
     134&    11038951&0.75&21.87&107.38$\pm$1.00&-&-&-&-&-&-20.30&-19.99&-19.89&-&-\\
     158&    11020790&0.93&22.08&795.07$\pm$163.23&-&-&-&-&-19.58&-&-&-&-&-\\
     165&    11014633&0.77&21.71&126.03$\pm$9.91&-&-&-&-&-19.02&  -19.63&  -19.35&  -18.46& 5.76&1.14\\
     172&    11015249&0.81&23.11&228.37$\pm$1.09&-&-&-&-&-&-&-&-&-&-\\
\label{table1}
\end{longtable}
\end{center}
\end{landscape}
\twocolumn

\section{Summary and Conclusions}
\label{sec:discussion}

We have investigated the nature of 61 70\,$\mu$m-selected sources, by
exploiting data from \textit{Spitzer}, \textit{Chandra} and the Keck
telescopes, the latter as part of the DEEP\,2 survey. In previous work
we showed that the energy budget of 70\,$\mu$m-selected galaxies is
defined by emission in the infrared, with all z$>$0.1
galaxies having luminosities above L$_{IR}$\,$>$\,10$^{10}$\,L$_{\odot}$. Our results
revealed that 70\,$\mu$m populations are predominantly comprised of
LIRGs at an average redshift of z$\sim$0.5 and star formation rates of
the order of $\sim$\,100\,M$_{\odot}$/yr. In addition, we found that
the strong majority are identified with a starburst-type SED: a
well-defined optical/near-IR stellar bump, followed by an inflection
at 2\,$\lesssim$ $\lambda_{rest}$ $\lesssim$ 6\,$\mu$m and a sharp
increase in flux at infrared wavelengths.  
The aim of the work described in this paper was to examine the X-ray
and spectral properties of these sources and, by building on previous
results, develop a more detailed picture of the inherent nature of
70\,$\mu$m-selected galaxies.

In order to place our sample in the context of other galaxy
populations in the DEEP\,2 photometric and spectroscopic survey, we
evaluated its spectral properties against a control sample assembled
from DEEP\,2 objects of similar redshifts and optical colours (section
\ref{sec:control_sample}). The clearly identifiable stellar bump in
our sources' SEDs, places them in the optically bright
(-18\,$<$\,M$_B$\,$<$\,-23) and red (0.5\,$<$\,U-B\,$<$\,1.5) regime
in the colour-magnitude diagram (CMD). 
Our results showed that although there is agreement between some
control and 70\,$\mu$m galaxies, there were also evident differences
both in terms of their AGN and stellar content. The [OIII]/H$\beta$
distributions of 70\,$\mu$m LIRGs and respective control galaxies were
in agreement, with only a few outliers, implying that the
optically-determined AGN fractions are broadly consistent in that
redshift and U-B/M$_{B}$ range. In contrast, there was a more evident
shift towards higher [OIII]/H$\beta$ for the ULIRGs than their
respective control galaxies. In terms of the [NeIII]/[OII] ratio half
of the 70\,$\mu$m sample was found to agree with the comparison
sample, whereas half was offset towards higher values. As discussed in
section \ref{sec:NeIII_OII}, high [NeIII]/[OII] ratios can be
representative of low metallicity systems, but also a harder
ionisation field, albeit not unambiguously. In terms of H$\delta$ EWs
and the H$\delta$-D$_n$4000 relation, the spread in values for the two
samples show little overlap. The 70\,$\mu$m population seems to be at
a different evolutionary state to the control sample, undergoing
intense starburst episodes and characterised by younger stellar
populations. This might not be surpising since, as mentioned earlier,
the place of the 70\,$\mu$m sample on the CMD diagram could have been
affected by extinction.

Our main results are:

\begin{itemize}

\item A total of 8 out of 61 sources (13 per cent of the sample) show at least one AGN
signature and are therefore classed as hosting an active nucleus. The AGN fraction is 0, 11
and 23 per cent for the SB, LIRG and ULIRG groups separately, robust for L$_{2-10keV}$\,$>$\,10$^{42}$\,erg/s and N$_H$\,$<$\,4$\times$10$^{23}$\,cm$^{-2}$, but potentially an overall lower limit due to a number of lower luminosity, more heavily obscured AGN that could be missed. [Note that by AGN fraction, we simply refer to the number of sources which host an
AGN]. The increase with total infrared luminosity is consistent with both local and high redshift studies
(e.g. Lutz et al. 1998\nocite{L98}, Fadda et al. 2002\nocite{F02}, Franceschini et al. 2003\nocite{F03}).

\item 7 out of the 8 sources which show signatures of an active nucleus, have SEDs which are starburst-type with the integrated infrared emission $\sim$\,1-3 orders of magnitude higher than the integrated AGN emission. One object has a power-law type SED, with a strong near/mid-IR continuum which implies that the AGN is energetically important. Obscured AGN which are possibly missed when estimating the AGN fraction, are not likely to play a role when estimating the fraction of AGN-dominated sources. We estimate that for an AGN to power a LIRG or ULIRG, it must have an X-ray luminosity of at least 10$^{43}$\,erg/s and up to 10$^{45}$\,erg/s. If the central black hole activity is powerful enough to contribute substantially or even dominate the galaxy's energy budget, unambiguous AGN signatures would emerge in the near/mid-IR part of the SED, even if missed in the X-rays. As a result, we conclude that all sources in the 70\,$\mu$m sample are primarily powered by star-formation --- this includes object 165 which has an energetically important AGN, as its far-IR SED is more luminous than the AGN component.

\item The X-ray detection fraction of the control sample is 4.2($\pm$1.6) per cent, whereas for the 70\,$\mu$m sources it is 16.4($\pm$6.4) per cent. In terms of the latter, about 50 per cent comes from the objects we identified as AGN, showing a strong link between the X-ray and AGN fractions. This is to be expected, as the EGS \emph{Chandra} survey is not deep enough to be sensitive to most of the star-forming sources in our sample. The higher X-ray, and hence, AGN incidence for the 70\,$\mu$m sample is at the $\sim$2$\sigma$ significance level. It possibly relates to the higher estimated average stellar mass, as black hole mass is expected to scale roughly with galaxy mass, however, it could also hint at a relation between AGN activity and star-formation (e.g. see also Kauffmann et al. 2003\nocite{Kauffmann03} and
Silverman et al. 2009\nocite{Silverman09} find a number of AGN that reside in infrared-selected star-forming hosts). It remains to be seen whether the higher AGN fraction in 70$\mu$m sources is due to the higher stellar mass or the more intense star-forming activity or both.

\item The AGN incidence of 13 per cent that we estimate is significantly lower than what has been
previously observed. Previous studies have shown that, although star-formation is the dominant energy source
in LIRGs and ULIRGs, a significant fraction of them host bolometrically important AGN (e.g.,
Solomon et al. 1997\nocite{Solomon97}; Veilleux, Sanders $\&$ Kim 1997\nocite{VSK97};
Genzel et al. 1998\nocite{Genzel98}; Downes $\&$ Solomon 1998\nocite{DS98};
Scoville et al. 2000\nocite{Scoville00}; Soifer et al. 2001\nocite{Soifer01};
Fadda et al. 2002\nocite{Fadda02}; Brand et al. 2006\nocite{Brand06}).
Given this discrepancy, one would naturally ask whether there is a bias with respect to the wavelength of selection. 
The current status of far-infrared astronomy implies that the majority
of studies on the AGN content of infrared galaxies have been based on
populations selected in the mid-IR with the MIPS 24\,$\mu$m band. As
discussed in section \ref{sec:agn}, there are fundamental differences
between AGN and starburst SEDs and although entirely AGN or starburst
dominated objects are the extremes of this range, objects with
AGN/starburst components of comparable contributions are rare and as a
result sources will fall in one of the two categories. Our results
have shown that a far-IR bump and an SED that peaks longward of
$\sim$50\,$\mu$m is strong evidence for the source being
starburst-dominated, as AGN-dominated sources peak at much shorter
(near/mid-IR) wavelengths and are weak far-IR emitters (see also
Alonso-Herrero et al. 2006\nocite{AH06a}). Consequently, one would
expect a higher AGN incidence in 24\,$\mu$m populations, as a) it
probes emission from warm/hot dust and b) at the mean redshift of the
population (z$\sim$0.8) it corresponds to
$\lambda_{rest}$\,$\sim$13\,$\mu$m, where there is little relevance to
ongoing star formation. 
It seems that setting the selection at 70$\mu$m, at least down to the
few mJy flux-density limit of our survey, enables the detection of a
population of almost entirely starburst-dominated galaxies. This is also
consistent with recent work by Trichas et
al. (2009)\nocite{Trichas09}; although the 24$\mu$m/X-ray initial
selection in Trichas et al. (2009) implies that most objects in their
sample host an energetically important AGN, the additional 70\,$\mu$m detection criterion
results in these systems also being strongly star-forming.

\end{itemize}

\section*{Acknowledgments}
This work is based on observations made with the Spitzer Space
Telescope, operated by the Jet Propulsion Laboratory, California
Institute of Technology, under NASA contract 1407 and partially supported by JPL/Caltech contract 1255094 to the University of Arizona. We thank the anonymous referee for valuable comments.

\bibliographystyle{mn2e}
\bibliography{references}

\label{lastpage}

\end{document}